\documentclass[traditabstract]{aa}

\usepackage[normalem]{ulem}

\usepackage{graphicx}
\usepackage{txfonts}
\usepackage{xcolor}
\usepackage{comment}
\usepackage{array}
\usepackage{natbib}
\bibpunct{(}{)}{;}{a}{}{,}
\usepackage{hyperref}
\hypersetup{colorlinks=true,
            citecolor=blue,
            linkcolor=blue}
\usepackage{color}
\newcommand{\orcid}[1]{\protect\href{https://orcid.org/#1}{\protect\includegraphics[width=8pt]{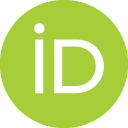}}}
\newcolumntype{H}{>{\setbox0=\hbox\bgroup}c<{\egroup}@{}}
\def\crm{\cr\noalign{\medskip}}
\def\vv#1{\vec{#1}}
\def\be {\begin{equation}}
\def\ee {\end{equation}}
\def\vr {\vv{r}}

\def\dvr{\vv{\dot r}}

\def\M{M}
\def\G{{\cal G}}
\def\H{{\cal H}}
\def\nug{\nu_1}
\def\nus{\nu_2}
\def\DeltaIn{\delta}

\def\figpath{}
\def\bibpath{}
\def \llabel#1{\label{#1}}
\begin{document}

\title{Dynamics of trans-Neptunian objects near the 3/1 mean-motion resonance with Neptune}

\titlerunning{Dynamics of the Threetinos}

\author{
Alan J. Alves-Carmo\inst{1} %\orcid{0000-0002-4651-2277}
\and
Timoth\'ee Vaillant\inst{1,2} %\orcid{0000-0002-1043-326X}
 \and
Alexandre C. M. Correia\inst{1,3} %\orcid{0000-0002-8946-8579} 
}
\authorrunning{A.J. Alves-Carmo, T. Vaillant \& A.C.M. Correia}

\institute{
CFisUC, Departamento de F\'isica, Universidade de Coimbra, 3004-516 Coimbra, Portugal
\and 
Instituto de Telecomunica\c c\~oes, Universidade de Aveiro, 3810-193 Aveiro, Portugal.
\and
IMCCE, UMR8028 CNRS, Observatoire de Paris, PSL Universit\'e, 77 Av. Denfert-Rochereau, 75014 Paris, France
}

\date{\today; Received; accepted To be inserted later}

  \abstract{
The complex classification of trans-Neptunian objects (TNOs) that are captured in mean-motion resonances (MMRs) and the constraint of their multiple origins are two significant open problems concerning the Solar System. The case-by-case study of the different MMRs and their characteristics provide information about their origin and dynamics, which helps us to understand the early stages of the Solar System evolution. In this paper, we study the dynamics of the detected TNOs close to a 3/1 MMR with Neptune. We initially use a semi-analytic three-body model to investigate the coplanar secular dynamics of these objects and find the stationary points. We then use surface sections and stability maps to analyse the non-averaged dynamics. These methods allow us to isolate the different stability regions and determine the extent of the chaotic regions. We show that stability maps are an extremely powerful tool for studying the resonant dynamics when they are computed in terms of the resonant angle. We then use these maps to study the non-planar three-body problem and the full dynamics in the presence of planetary perturbations. We confirm that TNOs near the 3/1 MMR regions can exist at very high inclinations. In the framework of the three-body problem, many of these objects can also be stable outside the 3/1 MMR owing to a Kozai secular resonance. However, when we take into account the perturbations of the four giant planets, the Kozai regions disappear and only the 3/1 MMR region remains, with eccentricities $e \lesssim 0.5$.
}


   \maketitle
%
%-------------------------------------------------------------------

\section{Introduction}

The physical mechanisms that lead to the presently observed characteristics of trans-Neptunian objects (TNOs) include phenomena that occurred in the past, such as the outward migration of Neptune initially proposed by \citet{Fernandez_Ip_1984}.
One of the most accepted scenarios is the Nice model \citep{Tsiganis_etal_2005, Gomes_etal_2005, Morbidelli_etal_2007}, where a great instability between the orbits of the giant planets scattered and ejected tens of thousands of planetesimals from their initial locations, which explains the modern-day absence of a dense TNO population. 
Important factors in the formation of the TNOs are the composition of the primordial disc;  
the locations where the migration of Neptune started and ended; and the migration speed.

Numerical simulations of an outward migration of Neptune in a primordial planetesimal Kuiper Belt composed of a cold-gas disc and particles with random semi-major axis between 20 and 80~au provide the capture of objects in a large number of mean-motion resonances (MMRs), with an overpopulation of the 2/1 and 3/2 MMRs \citep{Hahn_Malhotra_2005}.
Indeed, several TNOs are currently observed close to MMR with Neptune \citep[e.g.][]{Elliot_etal_2005, Lawler_etal_2019, Bannister_etal_2018, Crompvoets_etal_2022}.
These can be resonant, but they can also be scattering, detached, or classical objects. 
Distinguishing between these possibilities is a delicate task, but the outcome has important implications for understanding the composition and the past evolution of these bodies \citep{Gladman_etal_2008}. 
For instance, by comparing the results of five different Neptune migration models with the high pericentre distance of detected TNOs close to 5/2 and 3/1 MMRs, \citet{Lawler_etal_2019} showed that a grainy and slow Neptune migration provides the best match to the observational data.

The complexity of MMRs has been exhaustively studied. 
Many authors have concentrated their efforts on analysing the topology of the problem and the equilibrium points. 
For the planar case, equilibrium solutions have been found for many MMRs \citep[e.g.][]{Message_1958, Taylor_1983a, Taylor_1983b}.
Using a semi-analytical secular model, \citet{Beauge_1994} demonstrated the existence of asymmetrical equilibria for the 2/1 and 3/1 MMRs, but their absence for the 3/2 and 4/3 MMRs.
\citet{Voyatzis_etal_2018} determined the equilibrium points using periodic orbits, while \citet{Lan_Malhotra_2019}  studied them with Poincar\'e surface sections. 
For the non-planar case, equilibrium solutions were also obtained and studied \citep[e.g.][]{Ichtiaroglou_etal_1989, Hadjidemetriou_1993, Voyatzis_Kotoulas_2005, Gallardo_2019, Namouni_Morais_2020b}.
Finally, for some resonances, it has been shown that the influence of the remaining giant planets (Jupiter, Saturn and Uranus) does not significantly modify the position of the resonant equilibria in the parameter space \citep[e.g.][]{Saillenfest_Lari_2017, Saillenfest_etal_2017b, Malhotra_etal_2018, Lei_etal_2022}.
Nevertheless, perturbations from the giant planets can be important for the long-term evolution and capture inside the MMRs and for the formation of other TNO classes.

\citet{Hahn_Malhotra_2005} showed that a stirred-up Kuiper Belt, which should extend at least until $a=55$~au, is capable of capturing objects in the 3/1 MMR with Neptune.
\citet{Bannister_etal_2018} report seven TNOs captured in a 3/1 MMR with Neptune, while \citet{Crompvoets_etal_2022} confirm a total of 12 objects inside this resonance.
However, the objects with perihelion distances smaller than 36~au, which includes most of the TNOs close to the 3/1 MMR, are inside a chaotic region that provokes a diffusion in the orbital elements \citep{Duncan_etal_1995, Gladman_etal_2002, Fernandez_etal_2004}.
As a result, these bodies may be scattered and do not remain captured in resonance for a long period.
The main characteristics of their dynamics should then be determined to evaluate whether or not these bodies are really trapped in the 3/1 MMR.
Estimation of the amount of time spent by TNOs inside this resonance and determination of the mechanisms that lead these bodies to escape are also important, in order to compare the number of bodies expected in this region according to the current formation models with numbers derived from observations. 

In this paper, we focus our analysis on the TNOs that are found close to the $3/1$ MMR with Neptune, which are also known as {threetinos}.
In Sect.~\ref{minor_bodies}, we present a dynamical analysis of these objects carried out in order to ascertain whether or not they are locked in resonance.
In Sect.~\ref{semimodel}, we use a semi-analytic model to describe the secular planar dynamics around the 3/1 MMR and determine the stationary points of the problem. 
We then investigate how the secular model is modified in the non-averaged planar case using surface sections (Sect.~\ref{poincare_surface_section}) and stability maps (Sect.~\ref{stabmapsnum}). 
These maps are then used to investigate the non-planar dynamics and the perturbations from the remaining giant planets.
Finally, we discuss the main results obtained and our conclusions in Sect.~\ref{conclsec}.

\section{Dynamics of TNOs near the 3/1 MMR}
\label{minor_bodies}

In this section, we present a dynamical analysis of the already known TNOs close to a 3/1 MMR with Neptune. 
Using the IAU Minor Planet Center database\footnote{\label{fnote1} \href{https://www.minorplanetcenter.net/data}{https://www.minorplanetcenter.net/} (data from March 8, 2023).}, we identified 82 bodies with heliocentric semi-major axes between 61 and 64~au, which corresponds to orbital period ratios with Neptune of between 2.9 and 3.1. 
When converting to barycentric coordinates, the number of objects in this range drops to 76.

\begin{figure}
 \centering
   \includegraphics[width=0.95\columnwidth]{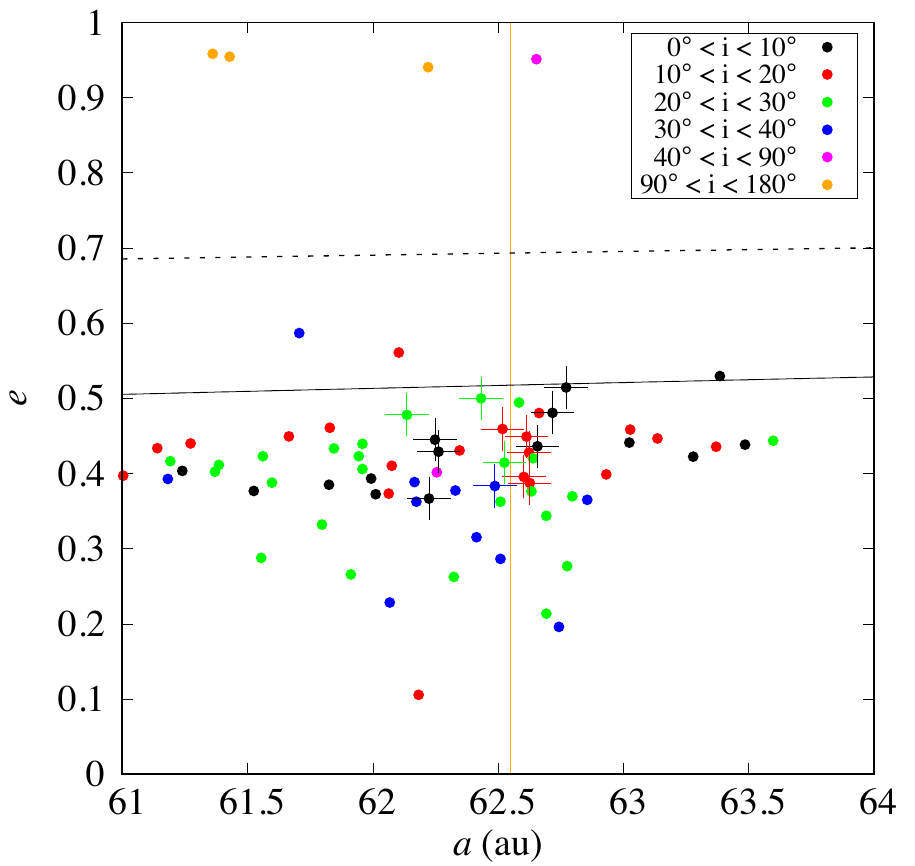}
      \caption{Barycentric orbital elements distribution of the TNOs that are close to the 3/1 MMR with Neptune. We show 76 bodies in the range $[61,64]$~au, with each colour corresponding to a different inclination range. The resonant TNOs are marked with a cross (Table~\ref{TabR}), and the remaining non-resonant with a dot. The orange vertical line gives the nominal resonant semi-major axis $a\approx 62.55$~au. The black solid and dashed lines give the overlap with Neptune and Uranus orbits, respectively.
 \label{classification_tno_13mmr}}
\end{figure}

In Fig.~\ref{classification_tno_13mmr}, we plot the 76 selected objects in a diagram showing semi-major axis versus eccentricity, where each colour signifies a different inclination range.
We note that most of the bodies have high eccentricities, in the range $0.3-0.5$, and have moderate and high inclinations of up to $40^\circ$.
Therefore, these objects are part of the scattering disc and detached populations. 
Seven objects have a sufficiently high eccentricity to cross the orbit of Neptune. 
Four of them, 2005\,{OE}, 2010\,{GW}$_{64}$, 2012\,{HD}$_2$, and 2013\,{HS}$_{150}$, have very eccentric orbits, very high inclinations, and a perihelion distance of $\sim 3$ and 10~au,
and as a result are not classified as TNOs; they belong to the Damocloid class objects \citep{Jewitt_2005}, and are not considered in our analysis. 

\subsection{Frequency analysis}
\llabel{freqanalys}

\begin{table*}
\tiny
\caption{List of TNOs in a 3/1 MMR with Neptune for at least 100~Myr.}    
\label{TabR}      
\begin{center}                         
\begin{tabular}{ l c c r r r r r r r r r}        
\hline\hline                 
TNO & Type & Resonant & Libration & Libration & Libration & $a$ & $e$ & $I$ & $\lambda$ & $\varpi$ & $\Omega$ \\    
& & angle &  centre (deg) & ampl. (deg) & period (kyr) & (au) & & (deg) & (deg) & (deg) & (deg)\\
\hline                        
$2000\,\textnormal{SS}_{331}$	&	S	&	 $\sigma_0$	&	78.4	&	40.4	&	19.1	&	62.771	&	0.514	&	2.984	&	272.899	&	272.990	&	146.809	\\
$2003\,\textnormal{LG}_{7}^*$	&	C	&	 $\sigma_0$	&	180.0	&	165.2	&	 $-$ 	&	62.135	&	0.478	&	20.104	&	1.740	&	342.235	&	238.346	\\
$2004\,\textnormal{VU}_{130}$	&	S	&	 $\sigma_0$	&	180.0	&	158.8	&	44.6	&	62.262	&	0.429	&	8.0236	&	325.186	&	271.547	&	42.958	\\
$2011\,\textnormal{US}_{411}^*$ &	S	&	all	&	70.5	&	31.1	&	18.2	&	62.431	&	0.500	&	22.041	&	17.679	&	9.271	&	24.813	\\
$2014\,\textnormal{FX}_{71}$	&	C	&	 $\sigma_0$	&	221.4	&	158.8	&	 $-$ 	&	62.517	&	0.459	&	17.726	&	225.553	&	193.391	&	275.858	\\
$2014\,\textnormal{SK}_{404}$	&	S	&	 $\sigma_0$	&	76.4	&	29.1	&	23.9	&	62.524	&	0.415	&	26.162	&	254.756	&	243.124	&	128.143	\\
$2014\,\textnormal{SR}_{373}$	&	S	&	 $\sigma_0$	&	71.7	&	22.4	&	27.9	&	62.486	&	0.383	&	35.583	&	232.502	&	214.479	&	165.746	\\
$2015\,\textnormal{GA}_{55}$	&	S	&	 $\sigma_0$	&	261.2	&	62.5	&	35.6	&	62.625	&	0.387	&	17.958	&	204.417	&	214.371	&	29.045	\\
$2015\,\textnormal{KY}_{173}^*$	&	S	&	 $\sigma_0$	&	273.5	&	44.2	&	24.4	&	62.656	&	0.436	&	8.996	&	119.677	&	135.324	&	152.811	\\
$2015\,\textnormal{RA}_{278}$	&	S	&	 $\sigma_0$	&	86.6	&	43.7	&	23.5	&	62.623	&	0.428	&	11.344	&	140.300	&	112.388	&	202.01	\\
$2015\,\textnormal{RZ}_{277}$	&	S	&	 $\sigma_0$	&	180.0	&	168.5	&	38.1	&	62.247	&	0.445	&	5.827	&	127.120	&	45.786	&	194.182	\\
$2015\,\textnormal{UH}_{87}$	&	C	&	 $\sigma_0$	&	157.9	&	155.1	&	 $-$ 	&	62.613	&	0.449	&	12.098	&	260.861	&	11.783	&	98.258	\\
$2015\,\textnormal{VN}_{166}$	&	S	&	 $\sigma_0$	&	76.4	&	31.4	&	20.2	&	62.716	&	0.481	&	6.030	&	146.592	&	119.125	&	227.357	\\
$2016\,\textnormal{SO}_{56}$	&	S	&	 $\sigma_0$	&	74.7	&	14.0	&	24.5	&	62.602	&	0.396	&	16.306	&	204.431	&	174.026	&	164.311	\\
$2019\,\textnormal{GZ}_{129}$	&	S	&	 $\sigma_0$	&	180.0	&	164.4	&	46.7	&	62.224	&	0.367	&	8.523	&	49.577	&	348.889	&	117.442	\\
\hline                              
\end{tabular} \\
\end{center}
TNOs that are shown in Fig.~\ref{stabfig} are marked with an asterisk.
The `S' stands for stable resonance (the libration frequency can be isolated from other perturbations), while `C' indicates that the libration angle displays strong chaotic behaviour.
We show the barycentric osculating orbital elements at the epoch JD2460000.5. 
\end{table*}

We first identified the bodies that can presently be captured inside the 3/1 MMR.
To this end, we numerically integrated all TNOs over 100~Myr, taking into account the perturbations from the four giant planets\footnote{The ephemerides of the giant planets at JD2460000.5 were taken from the JPL Horizons System (\href{https://ssd.jpl.nasa.gov/horizons/app.html}{https://ssd.jpl.nasa.gov/horizons})}.
We used the symplectic integrator SABA4 of \citet{Laskar_Robutel_2001} and an integration step size of 0.1~yr.
We then performed a frequency analysis of the heliocentric orbital elements \citep{Laskar_1990, Laskar_1993PD} to determine the fundamental frequencies ($n, \nug, \nus$) of the TNOs,  
where $n$ is the main frequency associated with the mean longitude (i.e. it is the average mean motion, which corresponds to the mean value of the mean motion over 100~Myr), and $\nug$ and $\nus$ are the main secular frequencies associated with the longitude of the perihelion and node, respectively.
Finally, we checked for combinations of these frequencies such that
\be
3 n - n' + (k-2) \nug - k \, \nus = 0 \ ,
\ee
where $n'$ is Neptune's average mean motion and $k$ is an integer.  
The exact equality almost never occurs numerically, and so we consider that the previous condition is satisfied whenever
\be
| 3 n - n' + (k-2) \nug - k \, \nus | / n' < 10^{-7} \ .
\ee
When such a combination exists, the associated resonant angle
\be
\sigma_k \equiv 3 \lambda - \lambda' + (k-2) \varpi - k \, \Omega 
\llabel{resonantangle}
\ee
is expected to librate around some equilibrium value,
where $\lambda$, $\varpi$, and $\Omega$ are the mean longitude, the longitude of the perihelion, and  the longitude of the node of the TNO, respectively, and $\lambda'$ is the mean longitude of Neptune.

In Table~\ref{TabR}, we list the 15 objects where it was possible to identify such a combination, all for $k=0$ (in Fig.~\ref{classification_tno_13mmr} we mark these objects with a cross).
We obtain the centre of libration, the libration period, and the amplitude of libration.
The frequency analysis also allows us to determine the additional forcing frequencies and amplitudes.
When the libration frequency can be clearly isolated from other perturbations, we classify the object as resonant stable (labelled with an `S' in Table~\ref{TabR}).
However, in some cases the libration angle displays strong chaotic behaviour, and the resonance may be unstable (labelled with a `C' in Table~\ref{TabR}).

Finally, we additionally identify the main frequency of $\varpi$ and $\Omega$.
In a decoupled problem, we expect that $\dot \varpi \approx \nug$ and $\dot \Omega \approx \nus$.
However, we observe that for one object, namely $2011\,\textnormal{US}_{411}$, both $\varpi$ and $\Omega$ precess with the same fundamental frequency.
As a result, the argument of the perihelion,
\be
\omega = \varpi - \Omega \ ,
\llabel{omegaangle}
\ee
also librates. 
In this case, all resonant angles $\sigma_{j\ne k}$ librate, because (Eq.\,(\ref{resonantangle}))
\be
\sigma_j = \sigma_k  + (j-k) \, \omega  \ .
\ee
Therefore, in Table~\ref{TabR}, we identify this case with the notation `all' for the resonant angle.

\citet{Crompvoets_etal_2022} report that objects $2013\,\textnormal{UB}_{17}$ and $2015\,\textnormal{VM}_{166}$ are also securely in a 3/1 MMR with Neptune, but these authors only integrate the orbits for 10~Myr.
These two TNOs are included in our selection of 76 objects near the 3/1 MMR, but we were not able to confirm their long-term resonant nature. 
Indeed, these objects are initially resonant, but after a few million years they display chaotic evolution. 
We use a different method to analyse the orbits and also integrate for a ten~times longer time-span.
As a result, we are able to detect the long-term chaotic diffusion of the orbits more precisely.
The initial orbital elements that we use may also slightly differ from those in previous studies, which can lead to a different orbital evolution.
These differences show that the classification of a TNO as a resonant object is not an easy task, and must be done with caution.

\begin{figure*}
\begin{center}
\includegraphics[width=1.0\textwidth]{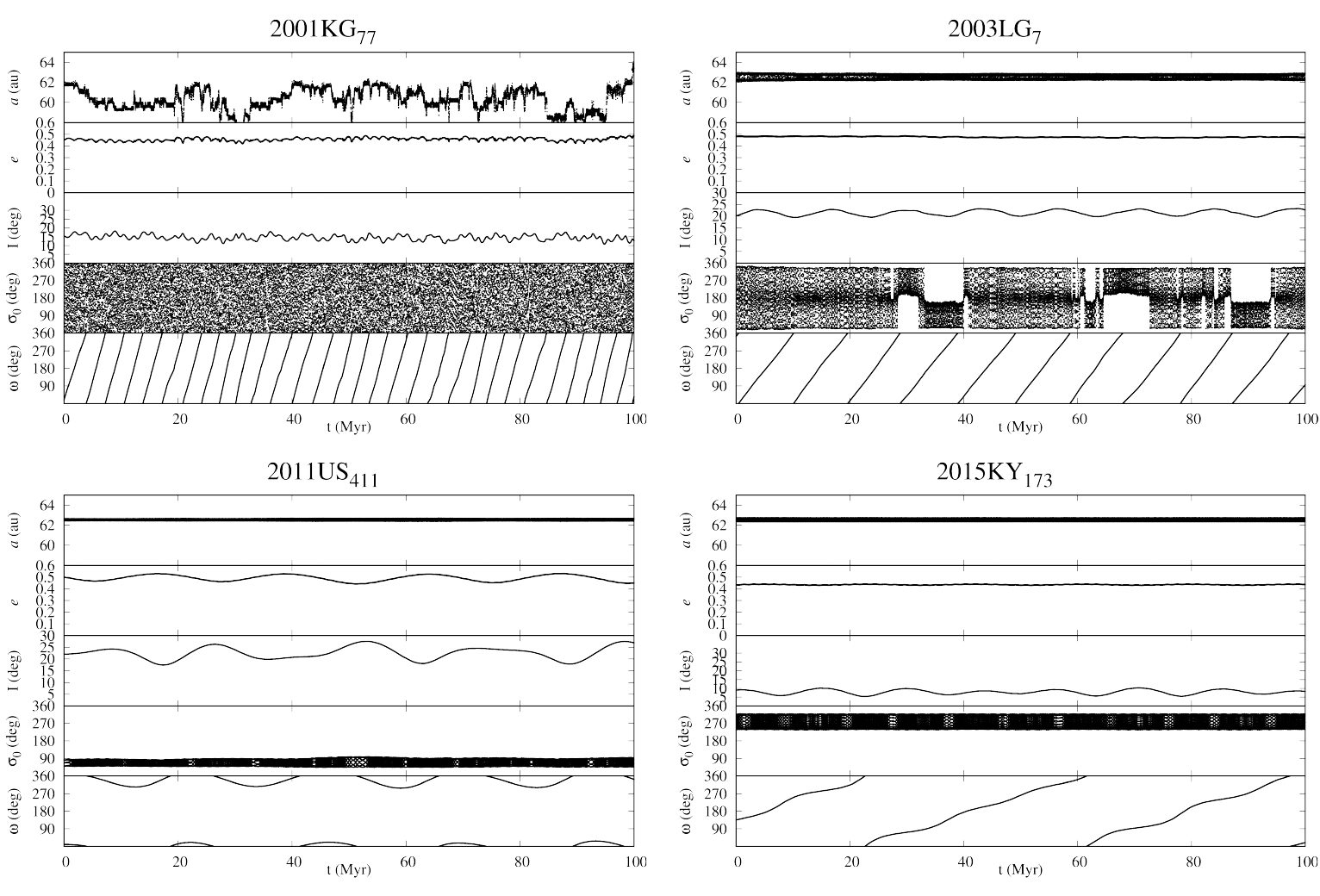}
\end{center}
\caption{Orbital evolution of some representative TNOs near the $3/1$ resonance over 100~Myr. We show the semi-major axis, the eccentricity, the inclination, the resonant angle $\sigma_0$, and the argument of the pericentre (from top to bottom). The initial conditions are taken from Table~\ref{TabR} and we take into account the perturbations from the four giant planets. We show the barycentric osculating orbital elements. \label{stabfig}}
\end{figure*}

In Fig.~\ref{stabfig}, we plot the orbital evolution of some TNOs near the $3/1$ MMR that better illustrate the diversity of the observed behaviours.
We first show an example of the many TNOs that are not captured in resonance, namely $2001\,\textnormal{KG}_{77}$.
We observe circulation for all resonant angles (Eq.\,(\ref{resonantangle})) and moderate chaotic evolution of the orbital elements.
We then show three examples of captured TNOs:
 $2011\,\textnormal{US}_{411}$ appears to be the most regular object in the family, as $\sigma_0$ and $\omega$ both librate with small amplitude;
$2015\,\textnormal{KY}_{173}$ has a nearly coplanar orbit and also a small libration amplitude, but only the angle $\sigma_0$ librates;
and finally, $2003\,\textnormal{LG}_{7}$ shows strong chaotic behaviour, although it remains in libration throughout the entire simulation. 
This latter object librates initially with large amplitude around $180^\circ$, but for some amount of time it librates around $100^\circ$ or $260^\circ$ with nearly half of the initial amplitude.

\subsection{Capture in resonance}
\llabel{migration}

Fifteen TNOs were identified in the 3/1 MMR with Neptune (Table~\ref{TabR}), clearly suggesting some past orbital evolution.
We therefore performed some quick numerical simulations of the outer migration of Neptune in the late stages of the evolution of the Solar System.
Our simulations start after the occurrence of the instability due to the crossing of the 2/1 MMR resonance between Jupiter and Saturn, when Neptune had already evolved into an outer orbit with respect to Uranus \citep{Tsiganis_etal_2005}.
We do not intend here to find the best migration model that reproduces the present distribution of bodies around the 3/1 MMR;
our goal is simply to understand the dynamical states acquired by the TNOs when they cross this resonance.
Therefore, we only include Neptune and the TNOs in the simulations.

\begin{figure}
\begin{center}
\includegraphics[width=0.95\columnwidth]{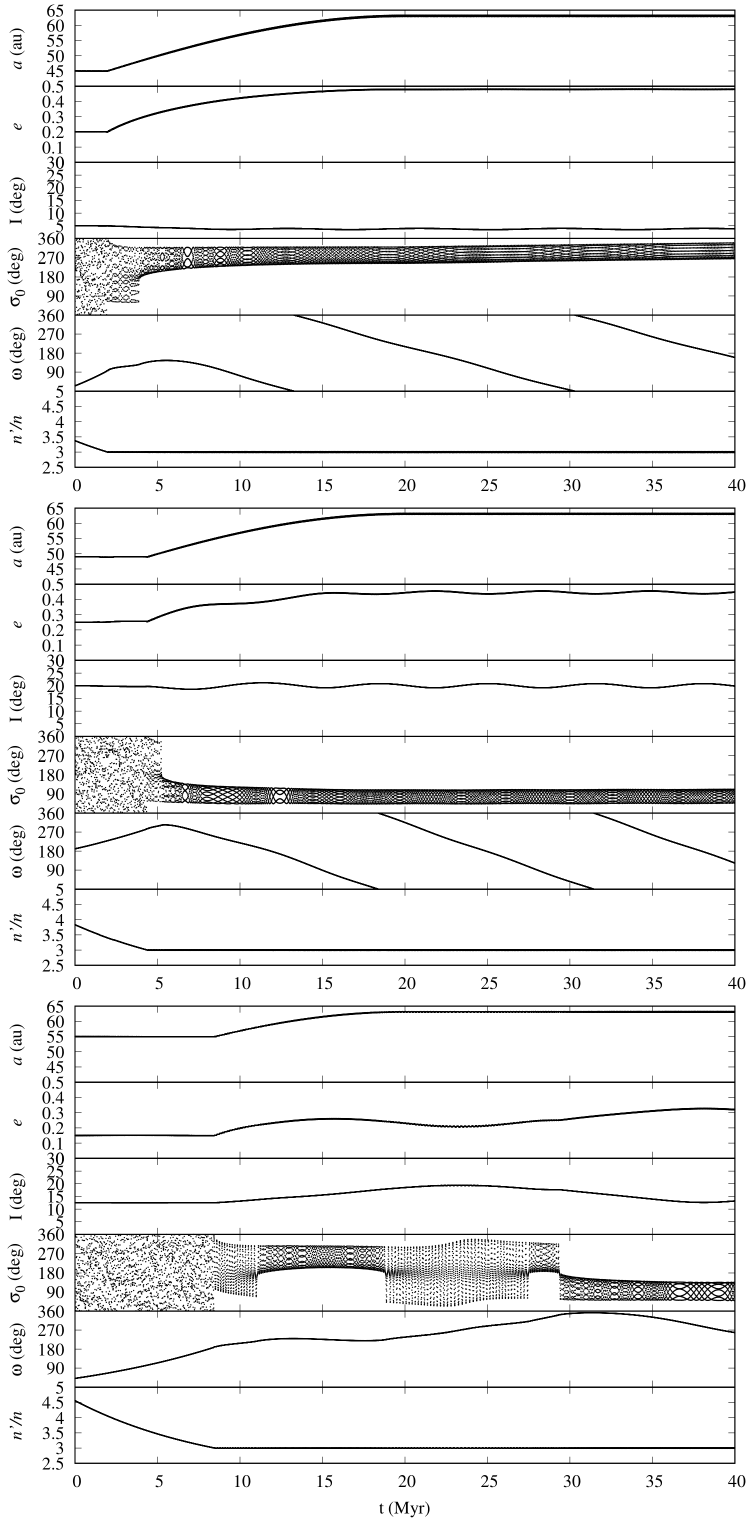}
\end{center}
\caption{Orbital evolution and capture in the 3/1 MMR of three TNOs during the migration of Neptune from 20 to 30~au. We show the semi-major axis, the eccentricity, the inclination, the resonant angle $\sigma_0$, the argument of the pericentre, and the ratio between the mean motion of Neptune and the TNO (from top to bottom).} 
\label{migration_neptune_capture_13mmr}
\end{figure}

Here, we adopt the migration model from \citet{Beauge_etal_2006}, which considers a Stoke's drag force acting on Neptune's orbit with an exponential decay of the semi-major axis with $\tau = 10^7$~yr.
We fix the initial value of Neptune's semi-major axis at 20~au, and vary the initial semi-major axes of the TNOs in the range 45$-$62~au with a step size of 1~au.
The initial eccentricities and inclinations are in the ranges $0-0.25$ and $0^\circ-25^\circ$, respectively, with step sizes of $0.025$ and $2.5^\circ$.
The distribution of the angles $\lambda$, $\varpi$, and $\Omega$ is random in the interval $[0^\circ, 360^\circ]$.
We therefore integrate a total of 2178 test particles over 40~Myr. 
Neptune reaches the present semi-major axis, $a\approx 30.15$~au, around 18~Myr.
Fig.~\ref{migration_neptune_capture_13mmr} shows some examples of TNOs captured in the 3/1 MMR.

\begin{figure}
\begin{center}
\includegraphics[width=0.95\columnwidth]{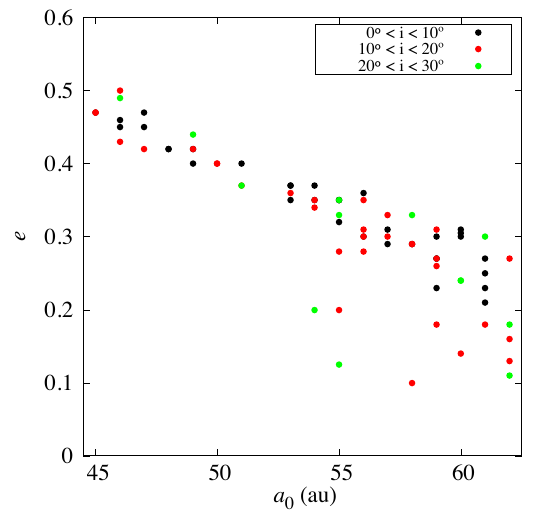}
\end{center}
\caption{Distribution of the final eccentricity of the TNOs captured in the 3/1 MMR as a function of the initial semi-major axis. We also show the initial inclination of these objects with colour classes. \label{eccdist}}
\end{figure}

We observe that 79 test particles were trapped in the 3/1 MMR at the end of the simulation ($\sim 4\%$).
Most of the captures occurred for the resonant angle $\sigma_0$, but 12 also involved the angle $\omega$, and 2 occurred for the angle $\sigma_2$ alone.
There were captures for all initial semi-major axis and inclination values, while for the initial eccentricity, capture was only observed when $e \ge 0.1$.
For the TNOs captured in resonance, the final semi-major axis ends around the present value of 62.5~au, and the initial inclinations remain nearly unchanged.
However, the initial eccentricity is always excited by the resonance and the final values range between 0.1 and 0.5.
Fig.~\ref{eccdist} shows the distribution of the final eccentricity of the TNOs captured in the 3/1 MMR as a function of the initial semi-major axis.
We observe that higher eccentricities are only possible for the TNOs with smaller initial semi-major axes, because they are trapped earlier and therefore the eccentricity has more time to grow.

From our simple numerical simulation, we conclude that the outward migration of Neptune can explain most of the presently observed dynamical properties of the {threetinos} (Table~\,\ref{TabR}).
The resonant angle is mainly $\sigma_0$, the final eccentricities are mostly within $0.1 < e < 0.5$, and the inclinations are random.
We also observe that, in some simulations, $\sigma_0$ librates with small amplitude around either $100^\circ$ or $260^\circ$, while in a few it librates with large chaotic amplitudes around $180^\circ$.
This last case is only observed for initial semi-major axes larger than 55~au.
An interesting feature present in most captured simulations is that before the libration centre settles at 
$100^\circ$ or $260^\circ$ with small amplitude, the TNOs initially librate around $180^\circ$ with large amplitude.
Therefore, the TNOs currently observed in this configuration were likely captured in the last stages of Neptune's migration.

\section{Secular dynamics}
\llabel{semimodel}

In this section, we describe the dynamics of the TNOs in the frame of the restricted three-body problem (Sun, Neptune, TNO), where the TNO is a test particle in an exterior orbit around Neptune.
The Hamiltonian of the system can be written as
 \be\label{hamiltonian1}
  \H =\H_k + \H_p = \dfrac{\left|\dvr\right|^2}{2}-\dfrac{\G \M}{\left|\vr\right|}-\G m' \left(\dfrac{1}{\Delta}-\dfrac{\vr\cdot \vr'}{\left|\vr^\prime\right|^3}\right) \ ,
 \ee
where $\G$ is the gravitational constant, $\M$ and $m'$ are the masses of the Sun and Neptune, respectively, $\vr$ and $\vr'$ are the positions of the TNO and Neptune with respect to the Sun, and $\Delta=\left|\vr-\vr^\prime\right|$.
The first two terms correspond to the Keplerian motion, $\H_k$, while the term in $m'$ corresponds to the interaction term, $\H_p$.

\subsection{Expansion in elliptical elements}

We first compute the resonant Hamiltonian by expanding $\H_p$ in elliptical elements.
We assume that the eccentricity of Neptune and its inclination to the reference plane are both zero ($e'=I'=0$).
For simplicity, we truncate the series up to the third order in the eccentricity of the TNO, $e$, and in the semi-major axis ratio, $\alpha = a' / a$, where $a$ and $a'$ are the semi-major axes of the TNO and Neptune, respectively.
We further perform the variable change of the mean longitudes $(\lambda, \lambda') \rightarrow (\sigma_0, \lambda')$, and average over the fast angle $\lambda'$.
We get\footnote{In the expression of $\cos \sigma_0$ we neglected the term in $\alpha^3$ because we already have terms in $\alpha^{-2}$.}
 \begin{eqnarray}\label{hamiltonian_res}
\langle \H_p \rangle &=& - \frac{\G m'}{a} \left[
 \frac{\alpha^2}{4} \left(1 + \frac32 e^2\right) \left( 1 - 6 s  + 6 s^2  \right) \right. \nonumber \\ & & 
- \frac38 \alpha^{-2} e^2  \left(1 - 9 \alpha^3\right)  \left(1-s \right) \cos \sigma_0 \nonumber \\ && 
\left. + \frac{15}{8} \alpha^3 \left(1-6 e^2\right) s \left(1-s\right)^2 \cos \sigma_2 \right] \ ,
 \end{eqnarray}
where $s = \sin^2 (I/2)$, and $I$ is the inclination of the TNO.
For the 3/1 MMR, we have $\alpha \approx 0.5$, and so we should have included more terms in the expansion of $\alpha$. 
For instance, terms in $\alpha^4$ include the secular contribution $\cos 2 \omega$ and the resonant angle $\cos \sigma_1$.
A complete expansion of the resonant Hamiltonian in terms of Laplace coefficients can be found in \citet{Namouni_Morais_2018b}.
However, our goal here is simply to identify the main resonant angles.
We see that for $I=0$, the only resonant angle is $\sigma_0$, while for $e=0$ the only resonant angle is $\sigma_2$.
We then search for a canonical change of variables that uses these two angles.

\subsection{Action-angle variables}

For the TNO, we first adopt the Delaunay canonical variables:
  \be
\begin{array}{l l l} 
L = \sqrt{{\cal G} \M a} \ , & G = L \sqrt{1-e^2} \ , & H = G \cos I  \crm
l = \lambda - \varpi \ , & g = \omega \ , & h = \Omega 
\end{array} \ . \label{delaunay_var}
  \ee
Neptune moves on an unperturbed orbit, with only one degree of freedom, given by $\lambda' = n' (t-t_0)$, where $n' = \sqrt{\G (\M+m') / a'^{3}}$, and $t$ is the time.
We assume the conjugate action of $\lambda'$ to be $\Lambda'$.
The Hamiltonian of the system becomes
\be\label{hamiltonian_delaunay}
\H = - \frac{\G^2 \M^2}{2 L^2} + n' \Lambda' + \H_p (L,G,H,l,g,h,\lambda') \ .
\ee
We introduce the new angles using a linear transformation
\be
\left[\begin{array}{c} 
\sigma_0 \\ \sigma_2 \\ \nu \\ \tilde \lambda
\end{array}\right] 
\equiv {\cal S} \,
\left[\begin{array}{c} 
l \\ g \\ h \\ \lambda'
\end{array}\right] \ ,
\ee
with
\be
{\cal S} =  \left[\begin{array}{cccc} 
3 & 1 & 1 & $-1$ \\
3 & 3 & 1 & $-1$ \\
3 & 3 & 3 & $-1$ \\
0 & 0 & 0 & $1$ 
\end{array}\right]  \ ,
\ee
which for the new actions gives \citep[e.g.][]{Goldstein_1950}
\be
\left[\begin{array}{c} 
S_0 \\ S_2 \\ N \\ \tilde \Lambda
\end{array}\right] 
= ( {\cal S}^{-1} )^T \,
\left[\begin{array}{c} 
L \\ G \\ H \\ \Lambda'
\end{array}\right] \ .
\ee
The new set of canonical variables that uses the resonant angles is then given by
\be
\begin{array}{l l} 
S_0 = \frac12 (L-G) \ , & \sigma_0 = 3 \lambda - \lambda' - 2 \varpi \crm
S_2 = \frac12 (G-H) \ , & \sigma_2 = 3 \lambda - \lambda' - 2 \, \Omega  \crm
N = \frac16 (3H-L)  \ , & \nu = 3 \lambda - \lambda'  \crm
\tilde \Lambda= \frac13 (3\Lambda' + L) \ , & \tilde \lambda = \lambda'  
\end{array} \ . \label{canonic_var}
\ee
The introduction of the angle $\nu$ is important, because this angle does not appear in the expression of the Hamiltonian for $e'=0$ (circular orbit for Neptune).
As a consequence, $N$ is a constant of motion (a circular orbit for the planet introduces an additional symmetry in the problem).
We further average over the fast angle $\tilde \lambda$ and finally get for the resonant Hamiltonian
\begin{eqnarray}\label{hamiltonian_canonical}
\langle \H \rangle &=& - \frac{\G^2 \M^2}{18 (S_0+S_2+N)^2} - n' (S_0+S_2+N) \crm
&& +  \bar \H_p (S_0,S_2,\sigma_0,\sigma_2; N) \ .
\end{eqnarray}
Here, $N$ is a parameter that depends on the initial conditions, and so the average resonant problem with $e'=0$ has only two degrees of freedom.

\subsection{Semi-analytical model}

The interaction term, $\bar \H_p$, can be obtained using series expansions, as in Eq.\,(\ref{hamiltonian_res}).
However, as the eccentricity of the threetinos is usually high (Table~\ref{TabR}), a large number of terms is required. 
An alternative and accurate way is to compute $\bar \H_p$ numerically over one orbital period of the TNO \citep[e.g.][]{Beauge_1994}:
\be\label{average_int}
\bar \H_p  =  -\G m'  \frac{1}{2\pi} \int_0^{2\pi} \left(\dfrac{1}{\Delta}-\dfrac{\vr\cdot \vr'}{\left|\vr^\prime\right|^3}\right) \, d \lambda \ .
\ee
The integration must be taken over $\lambda$ instead of $\lambda'$ to ensure that the system attains the initial configuration.
We also note that all orbital variables are fixed at the given values $(S_0,S_2,\sigma_0,\sigma_2; N)$ except $\lambda$ and $\lambda' = 3 \lambda - \sigma_0 - 2\varpi$. 
In order to numerically evaluate the integral (\ref{average_int}), it is more convenient to use the eccentric anomalies $E$ and $E'$, as the eccentricity of the TNO can reach high values \citep[e.g.][]{Pichierri_etal_2017}.
The position vector $\vr = (x,y,z)$ is given by 
  \begin{eqnarray} 
  \llabel{vrx}
  x & = & a\Big[\left(\cos{E}-e\right)\left(\cos{\omega}\cos{\Omega}-\sin{\Omega}\sin{\omega}\cos{I}\right) \\ & &- \sqrt{1-e^2} \sin{E}\left(\sin{\omega}\cos{\Omega}+\sin{\Omega}\cos{\omega}\cos{I}\right)\Big] \ ,  \crm \llabel{vry} 
  y & = & a\Big[\left(\cos{E}-e\right)\left(\cos{\omega}\sin{\Omega}+\cos{\Omega}\sin{\omega}\cos{I}\right) \\ & & - \sqrt{1-e^2}\sin{E}\left(\sin{\omega}\sin{\Omega}-\cos{\omega}\cos{\Omega}\cos{I}\right)\Big] \ , \crm \llabel{vrz} 
  z & = & a\Big[\left(\cos{E}-e\right)\sin{\omega}+\sqrt{1-e^2}\sin{E}\cos{\omega}\Big]\sin{I}  \ .
  \end{eqnarray}
The eccentric anomaly can be related with the mean longitude through the Kepler equation: $\lambda-\varpi=E-e\sin E$.
The average perturbation $\bar \H_p$ (Eq.\,(\ref{average_int})) is then computed as
\be\label{average_int2}
\bar \H_p  =  -\G m'  \frac{1}{2\pi} \int_0^{2\pi} \left(\dfrac{1}{\Delta}-\dfrac{\vr\cdot \vr'}{\left|\vr^\prime\right|^3}\right) (1-e\cos E) \, d E \ .
\ee

\subsection{Planar motion}
\label{planar_motion}

\begin{figure}
\centerline{\includegraphics[width=1\columnwidth]{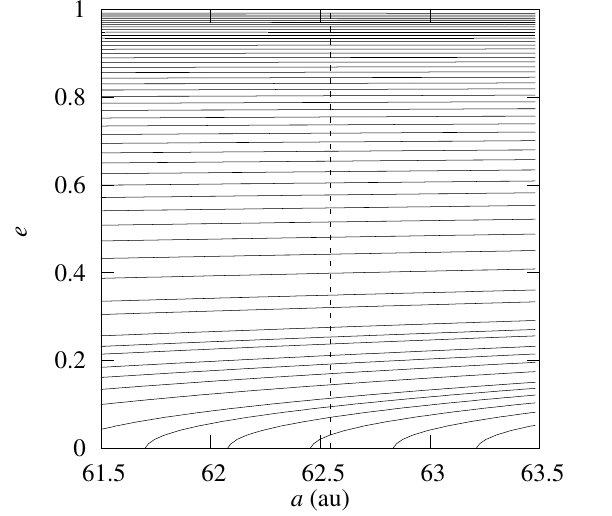}}
\caption{Level curves of the parameter $N$ in the $(a,e)$ plane (Eq.\,(\ref{Ne_value})). The vertical  line marks the position of the semi-major axis $a/a'=3^{2/3}$, which corresponds to the nominal 3/1 MMR.}
\label{n_curves_13MMR}
\end{figure}

The 3D restricted averaged resonant problem with $e'=0$ has two degrees of freedom (Eq.\,(\ref{hamiltonian_canonical})).
If we assume that the orbit of the TNO is on the same plane as Neptune, that is, $I=0$, we get $S_2 = (G-H)/2 = 0$ (Eq.\,(\ref{delaunay_var})).
The planar 2D problem therefore has only one degree of freedom, $\sigma_0$, and becomes integrable.
This allows us to understand the dynamics very easily from the level curves of the resonant Hamiltonian (\ref{hamiltonian_canonical}), which can be obtained using the semi-analytical approximation (Eq.\,(\ref{average_int})). 

In the planar case, the parameter $N$ simplifies as (Eq.\,(\ref{canonic_var}))
\be
N  =  \frac{1}{6} \sqrt{{\cal G} \M a} \left(3\sqrt{1-e^2} - 1\right) \ ; \label{Ne_value}
\ee
that is, the semi-major axis and the eccentricity are related.
In Fig.\,\ref{n_curves_13MMR}, we plot the level curves of constant $N$ in the $(a,e)$ plane for the 3/1 MMR.
We observe that for $e > 0.1$, the eccentricity is approximately constant for all semi-major axes in the vicinity of the resonance. 
For simplicity, we can parametrise the different types of Hamiltonians (Eq.\,(\ref{hamiltonian_canonical})) in terms of $e$ instead of $N$.
More precisely, in the following plots, we compute $N$ from expression (\ref{Ne_value}) using a given initial eccentricity and the nominal resonant semi-major axis $a=3^{2/3} a' \approx 62.55$~au.

\begin{figure}
\centerline{\includegraphics[width=1.0\linewidth]{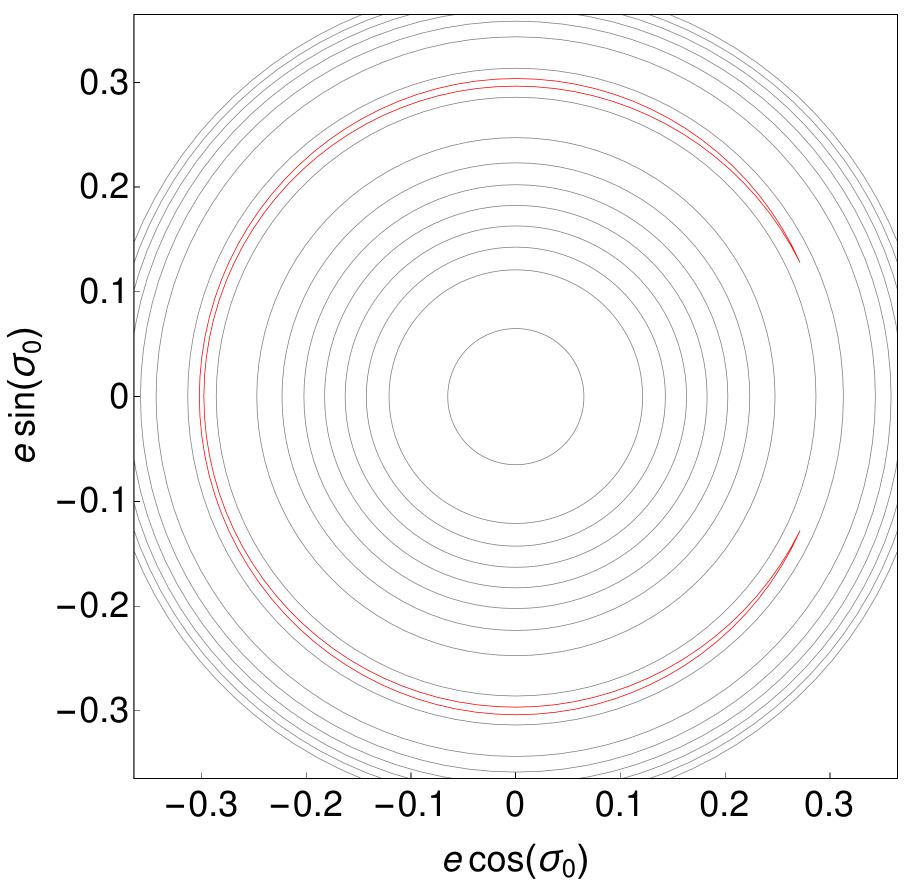}}
\caption{Level curves of the averaged Hamiltonian (\ref{hamiltonian_canonical}) in the plane ($e \cos \sigma_0, e \sin \sigma_0$) for $N = 15.43$~au$^2$/yr (corresponding to $e\approx0.3$ and $a/a' =3^{2/3}$). 
The contours of the narrow libration area (separatrix) corresponding to the 3/1 MMR is given in red.}
\label{level_curves_N_e03}
\end{figure}

Another consequence of $e$ being approximately constant for semi-major axes near the 3/1 MMR is a small amplitude for the eccentricity oscillations.
In Fig.~\ref{level_curves_N_e03}, we show the level curves of the Hamiltonian (\ref{hamiltonian_canonical}) for $e\approx0.3$ in the plane ($e \cos \sigma_0, e \sin \sigma_0$). 
We observe that the resonant island is indeed very narrow, which may explain why the capture probabilities in the 3/1 MMR in Sect.~\ref{migration} were relatively small ($\sim 4\%$).

\begin{figure*}
\includegraphics[width=1.0\textwidth]{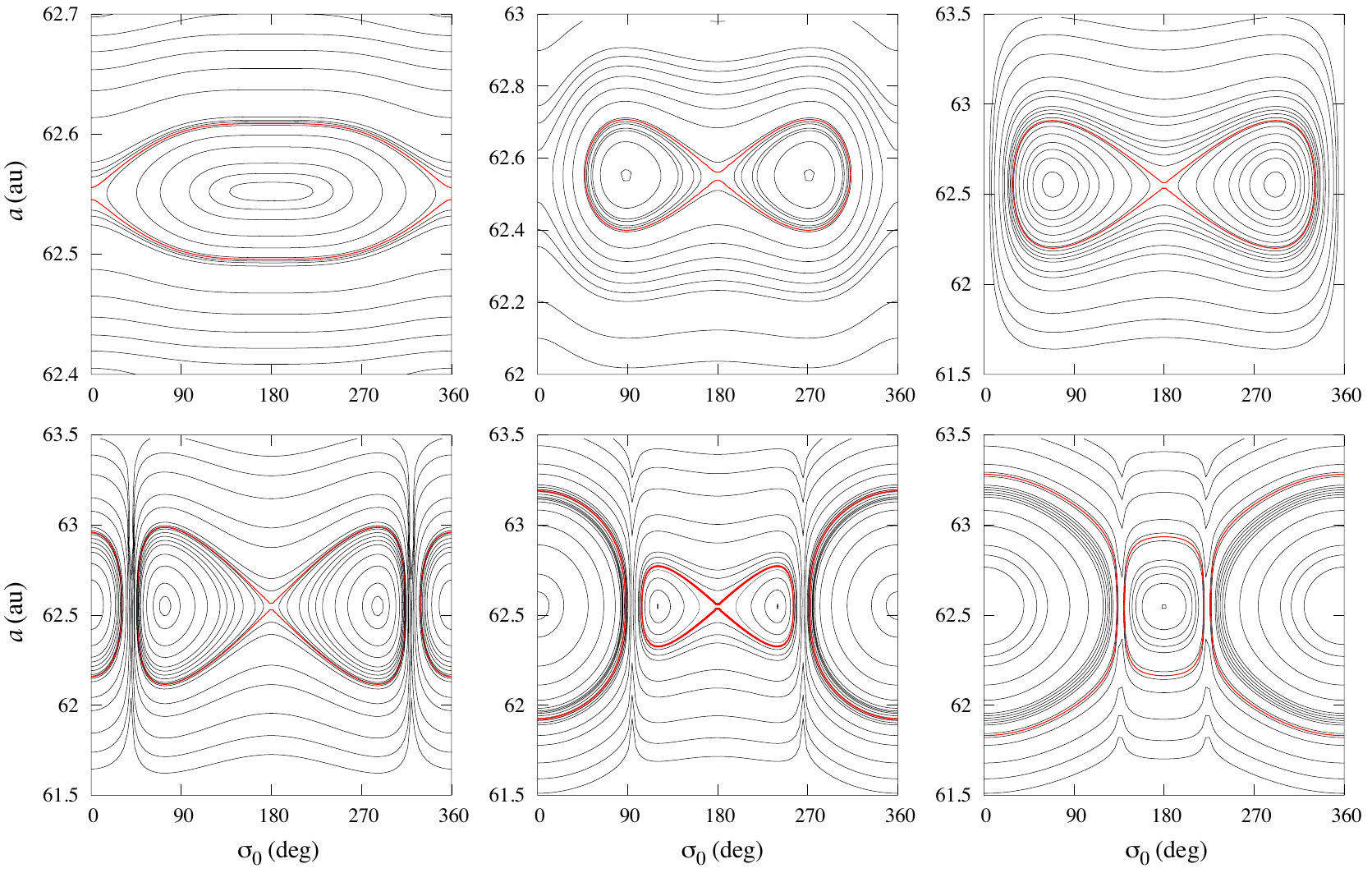}
  \caption{Level curves of the average Hamiltonian (\ref{hamiltonian_canonical}) for the 3/1 MMR in the plane ($\sigma_0, a$). For each panel, $N$ is constant and parametrised by $a/a' =3^{2/3}$ and the initial eccentricities (from left to right):  $e=0.1$, $e=0.3$, $e=0.5$ (top), and  $e=0.7$, $e=0.9$, $e=0.99$ (bottom). The separatrix of the resonant region is given in red. 
 \llabel{level_curves_13MMR}}
\end{figure*}

In order to better understand the topology of the 3/1 MMR, it is then more convenient to study the dynamics for the semi-major axis.
In Fig.~\ref{level_curves_13MMR}, we plot the level curves of the Hamiltonian (\ref{hamiltonian_canonical}) in the plane ($\sigma_0, a$) for different $N$-values corresponding to different initial eccentricities, from $e\approx0.1$ to $0.99$.
We can observe how the equilibrium points and resonance libration areas evolve with the eccentricity.
For $e < 0.13$, there is only a small libration zone around a stable equilibrium point centred at $\sigma_0 = 180^\circ$ (first resonance area).
There is also an unstable equilibrium point centred at $\sigma_0 = 0^\circ$.
For $0.13 < e < 0.52$, the equilibrium point at $\sigma_0 = 180^\circ$ becomes unstable, and two resonant areas appear centred at asymmetric $\sigma_0$ values, given by $\pm \sigma_r$.
A wider libration region that encircles the three previous equilibrium points is also present, which explains why in the migration simulations, just after being captured in resonance, $\sigma_0$ initially librates around $180^\circ$ before settling around $\pm \sigma_r$ with small amplitude (Fig.~\ref{migration_neptune_capture_13mmr}).
The width of the resonant areas increases with the eccentricity and the $\pm \sigma_r$ equilibria move away from $180^\circ$.
For $ 0.52 < e < 0.97$, the equilibrium point at $\sigma_0 = 0^\circ$ becomes stable, giving rise to a second resonance area.
As the eccentricity increases, the $\pm \sigma_r$ values approach $180^\circ$ again and the width of the first resonant area decreases, while the size of the second resonant area around $0^\circ$ increases.
Finally, for $e > 0.97$, the asymmetric equilibria at $\pm\sigma_r$ merge with the symmetric equilibrium at $\sigma_0 = 180^\circ$, which becomes stable again.

\begin{figure}
\centerline{\includegraphics[width=1.0\columnwidth]{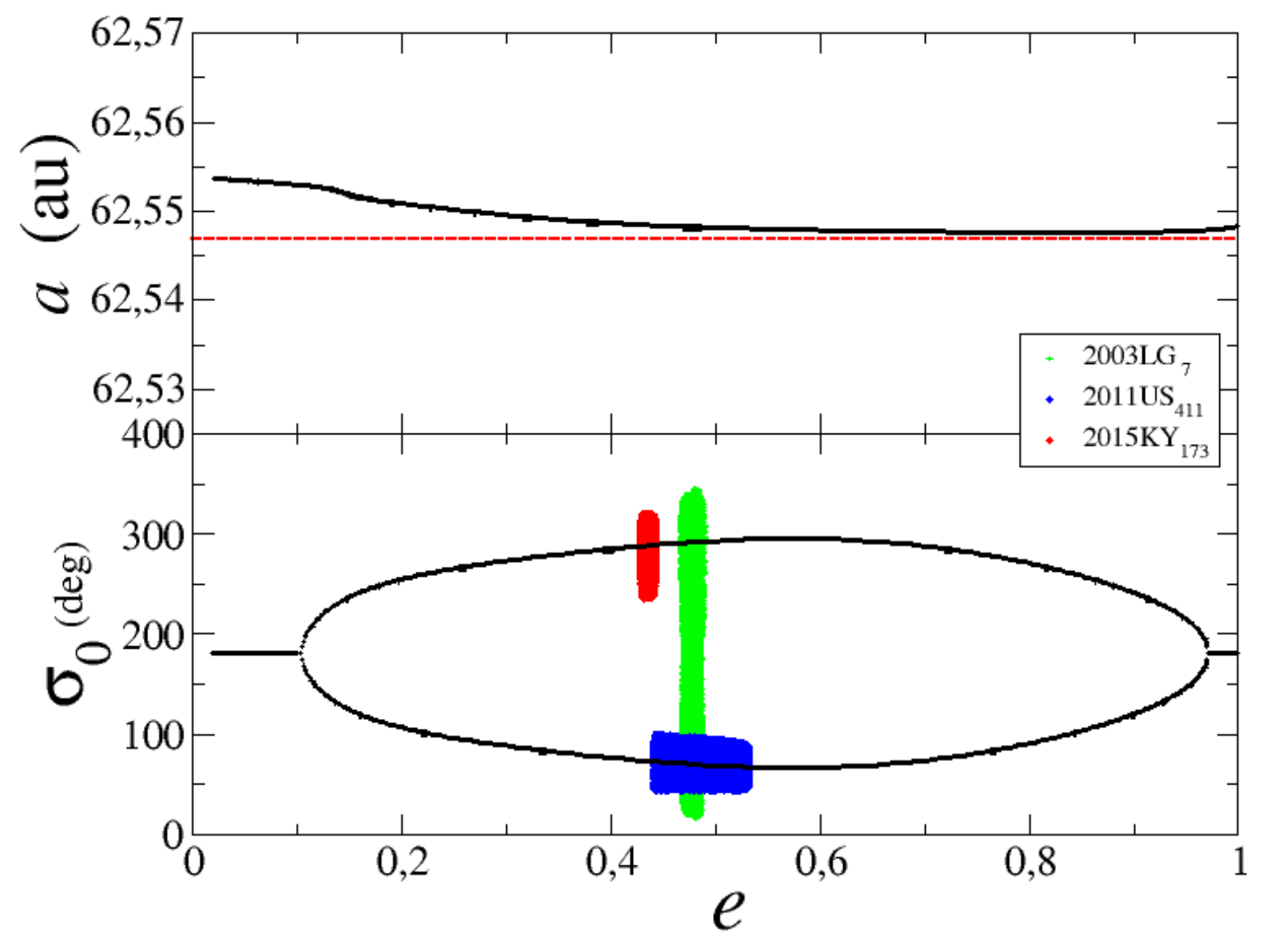}}
\caption{Equilibrium points for the 3/1 MMR with Neptune as a function of eccentricity. 
We show the equilibrium value for the semi-major axis (top) and the equilibrium value for the resonant angle (bottom). The red dashed line corresponds to the nominal value of the semi-major axis $a/a'=3^{2/3}$ for comparison.}
\label{aesigma_equilibrium_points}
\end{figure}

In Fig.~\ref{level_curves_13MMR}, we observe that the positions of the resonant equilibria $(\sigma_r, a_r)$ change with eccentricity;
they can be obtained by minimising the Hamiltonian (\ref{hamiltonian_canonical}),
\be
 \frac{\partial \langle  \H \rangle}{\partial a} = 0
 \quad \mathrm{and} \quad 
 \frac{\partial \langle  \H \rangle}{\partial \sigma_0} = 0  \ .
 \label{minHam}
\ee
For a given eccentricity, we first compute the corresponding $N$ value using expression (\ref{Ne_value}) with $a/a'=3^{2/3}$.
We then fix $\sigma_0=180^\circ$ and vary the semi-major axis to find the $a_r$ value that minimises the Hamiltonian.
We finally fix the semi-major axis at $a=a_r$ and vary $\sigma_0$ to find the $\sigma_r$ value that minimises the Hamiltonian.
In principle, the pair ($\sigma_0$ , $a$) that we find with this method could be away from the pair ($\sigma_r$ , $a_r$) that minimises the Hamiltonian (Eq.\,(\ref{minHam})), but we find that it is already a very good approximation \citep[see also][]{Pichierri_etal_2017}.
The results are shown in Fig.~\ref{aesigma_equilibrium_points}.
We observe that the $a_r$ value is always close to the Keplerian value given by $a/a'=3^{2/3}$.
Concerning $\sigma_r$, this latter moves away from $180^\circ$ as the eccentricity increases, with a maximum difference around $e=0.55$, for $\sigma_r \approx 60^\circ$.
For higher eccentricities, $\sigma_r$ approaches $180^\circ$ again.

Our results are in perfect agreement with those obtained by \citet{Voyatzis_etal_2018}, who used periodic orbits, and with those of \citet{Lan_Malhotra_2019}, who constructed Poincar\'e surfaces.
This shows that the semi-analytic secular Hamiltonian (\ref{average_int}) is also a good approach to represent the dynamics of the resonant three-body planar problem. 
Indeed, in Fig.~\ref{aesigma_equilibrium_points} we additionally show the results of the numerical integrations of the orbital evolution of three TNOs taken from Table~\ref{TabR}, whose evolution is also shown in Fig.~\ref{stabfig}.
We observe that the angle $\sigma_0$ of all these TNOs oscillates around one or both equilibrium values.

\section{Surface sections}
\llabel{poincare_surface_section}

\begin{figure*}
\includegraphics[width=1.0\textwidth]{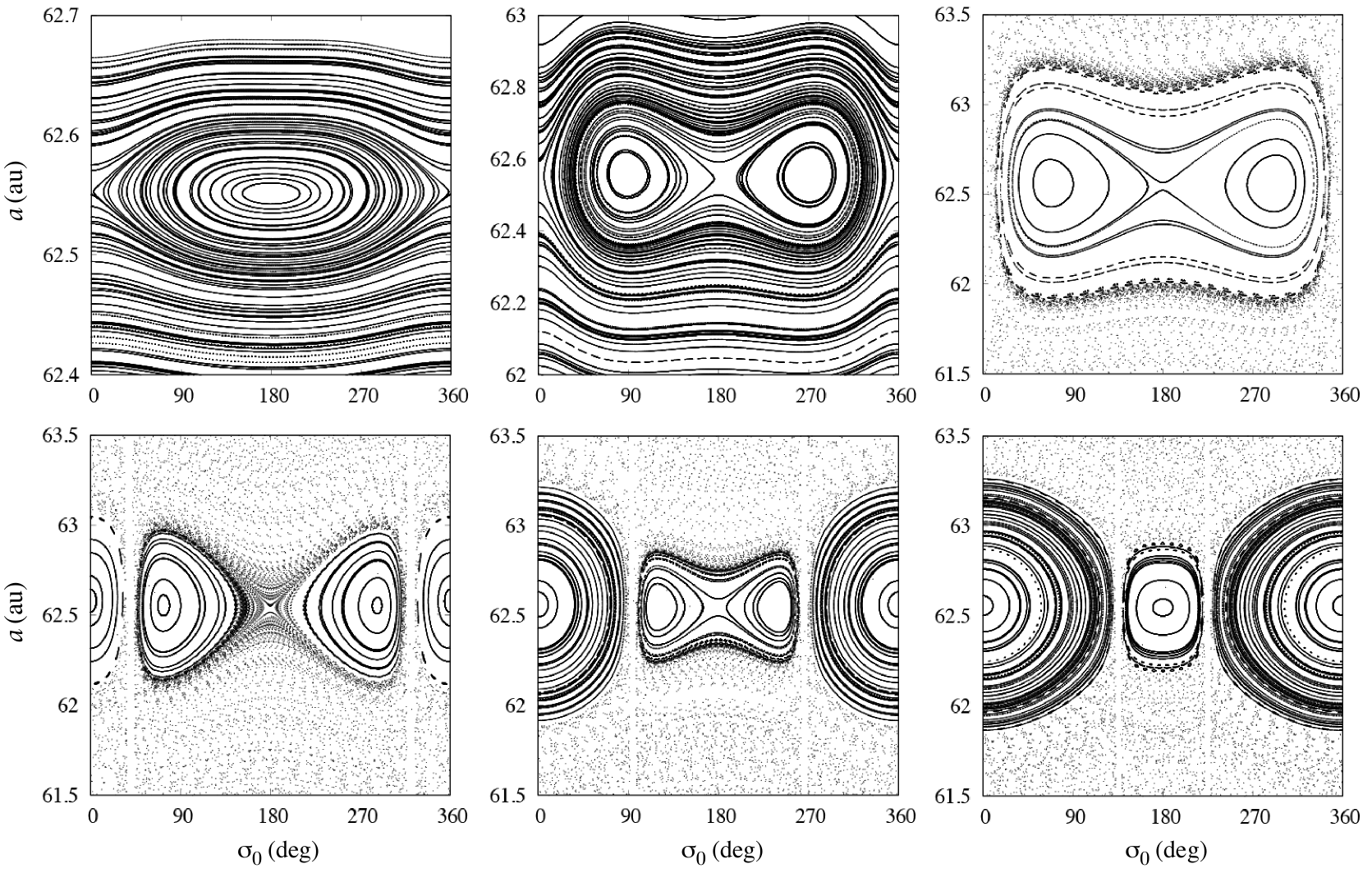}
      \caption{Surface sections in the plane ($\sigma_0, a$) for the three-body restricted planar problem close to the 3/1 MMR. For each panel, $N$ is constant and parametrised by $a/a' =3^{2/3}$ and the initial eccentricities (from left to right):  $e=0.1$, $e=0.3$,  $e=0.5$ (top), and  $e=0.7$, $e=0.9$, $e=0.99$ (bottom). \label{poincare_maps}}
\end{figure*}

In the previous section, we draw the level curves of the planar secular resonant Hamiltonian (Fig.~\ref{level_curves_13MMR}).
We observe that there are hyperbolic points and a separatrix surrounding the resonant equilibria, which can trigger chaotic motion in the non-averaged problem.
Therefore, following \citet{Lan_Malhotra_2019}, here we compute the Poincar\'e surface sections to get a more realistic picture of the planar dynamics for the 3/1 MMR.
The surface of section is crossed at the moment of the TNO passage of the pericentre, which corresponds to $\lambda=\varpi$ and $d \lambda / dt > 0$.

For the construction of the surface sections, we integrate the Newtonian three-body equations for a massless TNO for 2~Myr.
The surface sections are computed in the vicinity of the periodic solutions parametrised by $N$ (Eq.\,(\ref{Ne_value})), such that they can be compared with the level curves from Sect.~\ref{planar_motion}.
For a given $N$-value, the initial conditions are determined for different values of $a$ and $\sigma_0$.
However, in this case, we need to start the simulations with a set of osculating coordinates.
To get a better comparison with the secular problem (Eq.\,(\ref{hamiltonian_canonical})), we additionally use a {finite response filter} \citep{Carpino_etal_1987}.
This procedure consists in digitally filtering the output of the numerical integration by removing the fast frequencies with corresponding periods below 480 years.
This choice is sufficient to obtain similar curves to the approximation using the semi-analytic model (Fig.~\ref{level_curves_13MMR}).

In Fig.~\ref{poincare_maps}, we show the surface sections in the plane ($\sigma_0, a$) for the same $N$-values as in Fig.~\ref{level_curves_13MMR}, corresponding to different eccentricities, from $e \approx 0.1$ to $0.99$.
We observe that the surface sections provide similar results to the semi-analytic secular model.
However, as expected, there is one important difference: as the eccentricity increases, the trajectories near the separatrix become chaotic.
As a consequence, for $e \ge 0.7$, stable motion is only possible around one of the equilibrium points.
For smaller values, we can still have stable trajectories around both equilibria.
However, the presence of moderate chaos near the separatrix can trigger temporary excursions to different libration centres, as previously observed for some TNOs (Fig.\,\ref{stabfig}) and in the migration simulations (Fig.\,\ref{migration_neptune_capture_13mmr}).

\section{Stability maps}
\llabel{stabmapsnum}

\begin{figure*}
\includegraphics[width=1.0\textwidth]{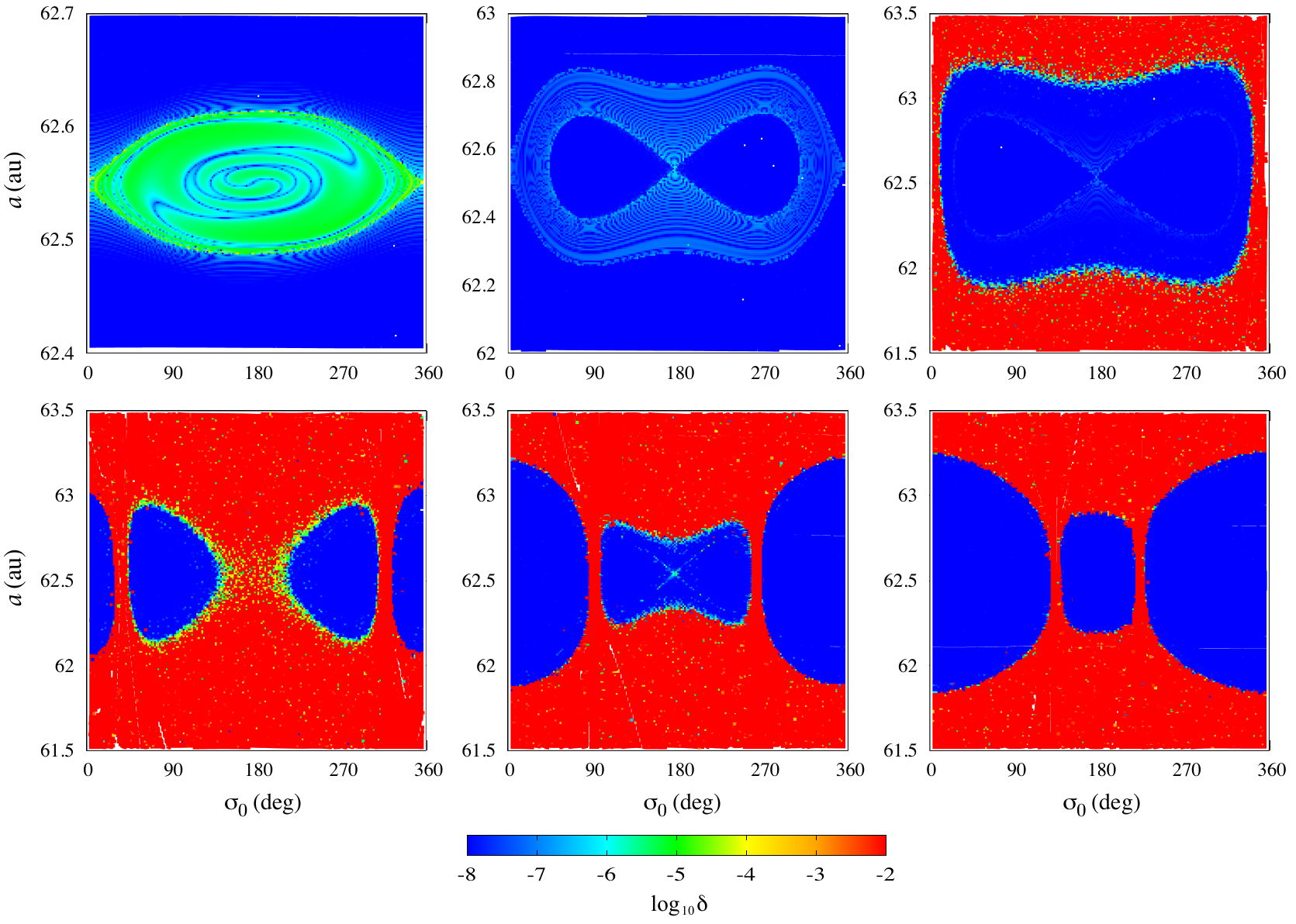}
      \caption{Stability maps in the plane ($\sigma_0, a$) for the three-body restricted planar problem close to the 3/1 MMR. For each panel, $N$ is constant and parametrised by $a/a' =3^{2/3}$ and the initial eccentricities (from left to right):  $e=0.1$, $e=0.3$,  $e=0.5$ (top), and  $e=0.7$, $e=0.9$, $e=0.99$ (bottom). The colour scale corresponds to the relative frequency diffusion index in logarithmic scale (Eq.\,(\ref{deltaindex})). \label{stabmap_planar_sigma0_filter_scaled}}
\end{figure*}

The semi-analytic secular model (Sect.\,\ref{semimodel}) provides a good determination of the stationary points of the Hamiltonian, but the correct dynamics in the 3/1 MMR can only be accessed using a complete three-body model, such as the one obtained with the surface sections (Sect.\,\ref{poincare_surface_section}).
Above, we restrict our analysis to the plane, although most {threetinos} lie on inclined orbits (Table~\ref{TabR}).
Surface sections are very useful to describe problems with two degrees of freedom (such as the planar non-averaged problem), but their use in studying problems with three or more is less straightforward.
Therefore, to study the global dynamics, in this section we adopt a method based on stability maps.
To this end, we use the frequency analysis method \citep{Laskar_1990, Laskar_1993PD} to map the diffusion of the orbital evolution of the TNO.

We consider a grid of initial conditions and integrate the Newtonian three-body equations of motion for a time $T$.
We then perform a frequency analysis of the mean longitude of the TNO, $\lambda$, using the software TRIP \citep{Gastineau_Laskar_2011} over the time intervals $[0,T/2]$ and $[T/2,T]$, and determine the main frequency in each interval, $f_1$ and $f_2$, respectively.
The stability of the orbit is measured by the index
\begin{equation}
\DeltaIn \equiv \left\lvert1 - \frac{f_2}{f_1}\right\rvert \ ,
\llabel{deltaindex}
\end{equation}
which estimates the stability of the mean motion long-distance diffusion \citep{Dumas_Laskar_1993}.
The larger $\DeltaIn$, the more unstable the orbital motion of the TNO.
For stable motion, we have $\DeltaIn \sim 0$, while $\DeltaIn \ll 1$ if the motion is weakly perturbed, and $\DeltaIn \sim 1$ when the motion is irregular.
It is difficult to determine the precise value of $\DeltaIn$ for which the motion is stable or unstable, but a threshold of stability $\DeltaIn_s$ can be estimated such that most of the trajectories with $\DeltaIn < \DeltaIn_s$ are stable \citep[for more details see][]{Couetdic_etal_2010}.

The diffusion index depends on the considered time interval.
Here, we integrate the equations of motion for $T=2$~Myr, because this interval is able to capture the main characteristics of the dynamics regarding the resonant frequency, which lies within the range $10^4-10^5$~yr (Table~\ref{TabR}).
With this time interval, we estimate that $\DeltaIn_s \sim 10^{-4}$.
The diffusion index $\DeltaIn$ is represented by a logarithmic colour scale calibrated such that blue and green correspond to stable trajectories ($\DeltaIn \ll \DeltaIn_s$), while orange and red correspond to chaotic motion ($\DeltaIn \gg \DeltaIn_s$).
Finally, as for the surface sections (Sect.~\ref{poincare_surface_section}), 
we apply the {finite response filter} \citep{Carpino_etal_1987}, to cut off the fast frequencies with periods below 480 years, which corresponds to a numerical average over the fast-frequency terms of the Hamiltonian (Eq.\,(\ref{hamiltonian1})).
However, in this case, the filter is only used to correct the values of the initial osculating orbital elements. 
The diffusion index (Eq.\,(\ref{deltaindex})) is still obtained with the frequency analysis of the full integration, because the main frequency of $\lambda$ is the mean motion (a fast frequency).

\subsection{Planar case}
\llabel{plancase}

We plot the stability maps in the plane ($\sigma_0, a$) using a grid of initial conditions with $200\times200$ points.
As in the previous sections, for each grid, we fix the value of $N$. 
Then, for each grid point ($\sigma_0, a$), we compute the eccentricity from $N$ using expression (\ref{Ne_value}).
In the restricted planar problem with $e'=0$, the dynamics does not depend on the angle $\nu$ (Eq.\,(\ref{hamiltonian_canonical})), and therefore we put $\nu=0$.
The initial position of Neptune is also irrelevant, and so we additionally set $\lambda'=0$, which implies that $\lambda = (\nu + \lambda')/3 = 0$.
For the longitude of the pericentre, we have $\varpi = (3\lambda-\lambda'-\sigma_0)/2 = -\sigma_0/2$.
Using expressions (\ref{vrx})$-$(\ref{vrz}), we finally get the initial condition for each grid point to use in a Newtonian three-body code.

In Fig.~\ref{stabmap_planar_sigma0_filter_scaled}, we show the stability maps in the plane ($\sigma_0, a$) for the same $N$-values as in Figs.~\ref{level_curves_13MMR} and~\ref{poincare_maps}, corresponding to different eccentricities, from $e \approx 0.1$ to $0.99$.
We observe that the stability maps provide similar results to the semi-analytic secular model (Fig.~\ref{level_curves_13MMR}) and to the surface sections (Fig.~\ref{poincare_maps}).
We are no longer able to determine the exact position of the stationary points of the problem, 
but the different dynamical regimes of interest become quite visible, in particular the distinction between stable (blue and green) and unstable regions (red).

We observe that the circulation regions outside the 3/1 MMR are completely stable for $e \le 0.3$, but they become unstable for $e \ge 0.5$.
Indeed, TNOs located around 62.5~au with  $e \ge 0.5$ will cross the orbit of Neptune (Fig.~\ref{classification_tno_13mmr}), and can therefore only be stable if protected by the 3/1 MMR.
Interestingly, the resonant areas that encircle a single asymmetric equilibrium point $\pm \sigma_r$ or both are also distinguishable.
For $e \le 0.5$, these regions are both stable, but in the first case there is less diffusion (dark blue regions) than in the second case (light blue regions).
This explains why the libration amplitude of the TNOs in this region presents some chaotic behaviour (Figs.~\ref{stabfig} and~\ref{migration_neptune_capture_13mmr}).

For $e=0.1$, we also note that the libration resonant region presents a higher level of diffusion than the surrounding circulation zone. 
This may explain why in the simulations with the migration of Neptune (Sect.~\ref{migration}) we never observed any capture in resonance for TNOs with $e<0.1$.
On the other hand, for $e > 0.7$, we observe that the second resonant region emerging around $\sigma = 0^\circ$ is also completely stable.

We conclude that the stability map method presented in this section gives similar results to the more traditionally used methods presented in Sects.~\ref{semimodel} and~\ref{poincare_surface_section}.
However, we believe the former has some practical advantages: it is easier to implement than semi-analytical approximations or surface sections; it clearly distinguishes stability regions from the unstable ones and quantifies the orbital diffusion; and, more importantly, it is not limited by the number of degrees of freedom in the problem.
As a result, it can be used to study the non-planar dynamics (Sect.~\ref{nonplanardyn}), or the dynamics in the presence of the remaining planets (Sect.~\ref{otherplanets}).

\subsection{Non-planar case}
\llabel{nonplanardyn}

We now consider that the inclination of the TNOs is no longer zero ($I\ne0$), as many of these objects are observed with moderate and high inclinations (Fig.~\ref{classification_tno_13mmr}).
In this case, the secular problem has two degrees of freedom (Sect.~\ref{semimodel}), which prevents us from building an integrable model.
It is also difficult to study the secular dynamics, because given the high eccentricity of the TNOs (Fig.~\ref{classification_tno_13mmr}), we need to include a large number of terms in the expansion of the interacting Hamiltonian (Eq.\,(\ref{hamiltonian_res})).
This is why in the planar model we used a semi-analytical approach to accurately compute the secular perturbations.
As explained in Sect.~\ref{plancase}, stability maps are adapted to study the planar secular dynamics, and so here we adopt the same approach to study the non-planar case.

For a better comparison with the planar case, we again plot the stability maps in the plane ($\sigma_0, a$) using a grid of initial conditions with $200\times200$ points.
Neptune is on a circular orbit and lies in the plane of reference.
In the restricted non-planar problem with $e'=0$, the dynamics does not depend on the angle $\nu$ (Eq.\,(\ref{hamiltonian_canonical})), and we can set $\nu=0$ and $N$ as constants.
Therefore, for each grid we still fix the value of $N$, which is now given by (Eq.\,(\ref{canonic_var}))
\be
N =  \frac16 \sqrt{{\cal G} \M a} \left(3\sqrt{1-e^2} \cos I -1\right)  \ ;
\llabel{N_nonplanar}
\ee
that is, it depends also on the inclination.
$N$ is parametrised for the nominal semi-major axis $a/a'=3^{2/3}$, and using a given initial eccentricity and inclination of the TNO.
Then, for each grid point ($\sigma_0, a$), we compute the eccentricity from $N$ using expression (\ref{N_nonplanar}) for the same initial value of the inclination.
This choice can be contested, because for each $N$-value we have different pairs $(e,I)$ that correspond to the same grid point ($\sigma_0, a$).
However, as we recalculate the eccentricity in the planar case, we continue in this way in order to be able to make a comparison.
Moreover, this is not a real problem, because as we use a three-body integration to produce the stability maps, the inclination is free to vary.
Concerning the remaining orbital parameters, we set $\lambda'=0$, as in the planar case.
As $\nu=0$, we still have $\lambda = (\nu + \lambda')/3 = 0$, and for the longitude of the pericentre, $\varpi = (3\lambda-\lambda'-\sigma_0)/2 = -\sigma_0/2$.
However, the choice of $\Omega$ is not irrelevant, because we now have one additional degree of freedom. 
For simplicity, we initially chose $\Omega = 0^\circ$, but we also explore different initial values for this parameter.
The initial condition for each grid point to use in the Newtonian three-body code is then finally obtained using expressions (\ref{vrx})$-$(\ref{vrz}).

\begin{figure*}
\includegraphics[width=1.0\textwidth]{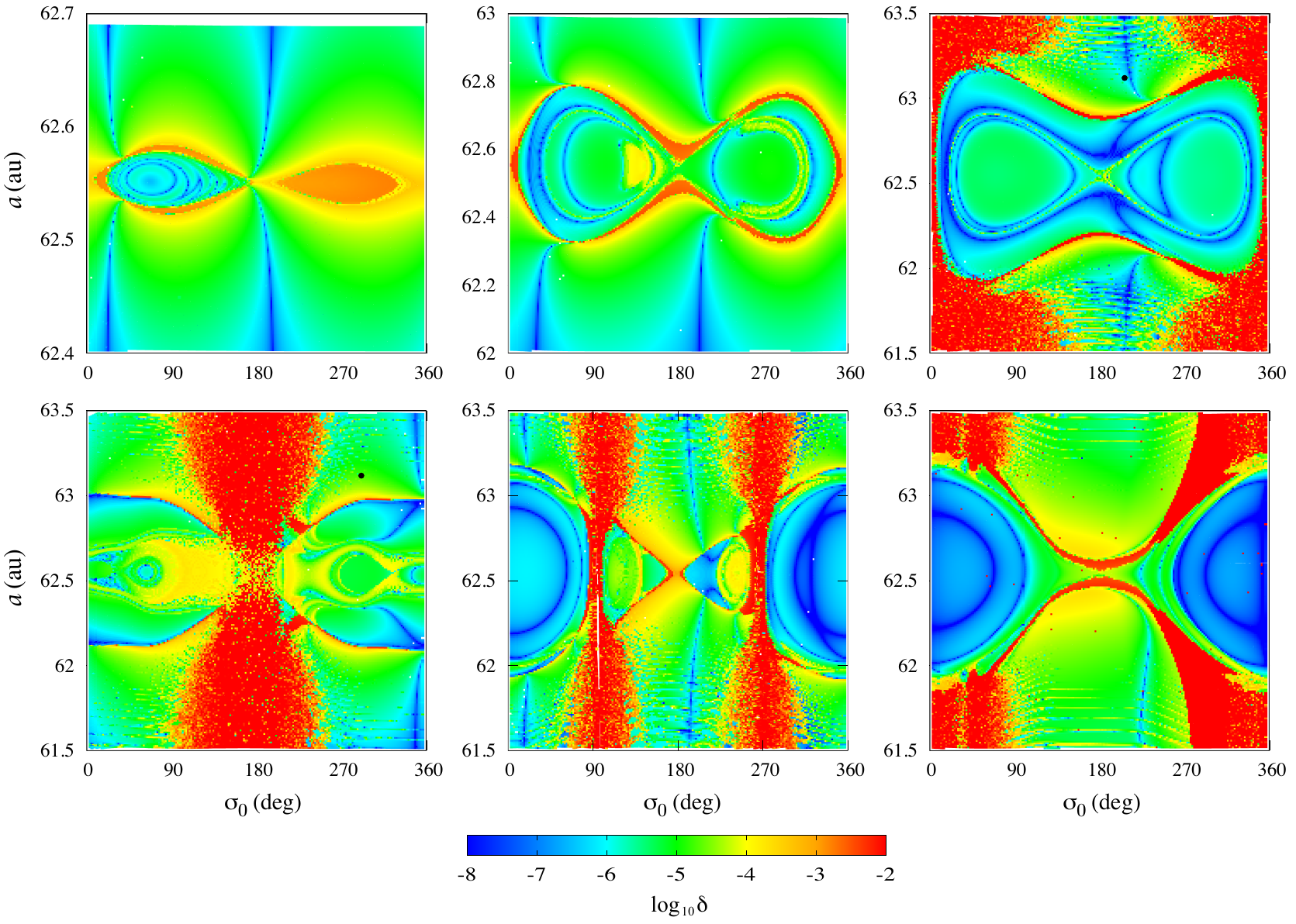}
      \caption{Stability maps in the plane ($\sigma_0, a$) for the three-body restricted problem close to the 3/1 MMR with $I=30^\circ$ and $\Omega=0^\circ$. For each panel, $N$ is constant and parametrised by the initial eccentricities (from left to right):  $e=0.1$, $e=0.3$,  $e=0.5$ (top), and  $e=0.7$, $e=0.9$, $e=0.99$ (bottom). The colour scale corresponds to the relative frequency diffusion index in logarithmic scale (Eq.\,(\ref{deltaindex})). The black dots correspond to the initial conditions whose long-term evolution is shown in more detail in Fig.~\ref{kozaitime}. \label{stabmap_i30_sigma0_filter_scaled}}
\end{figure*}

\begin{figure}
\begin{center}
\includegraphics[width=0.95\columnwidth]{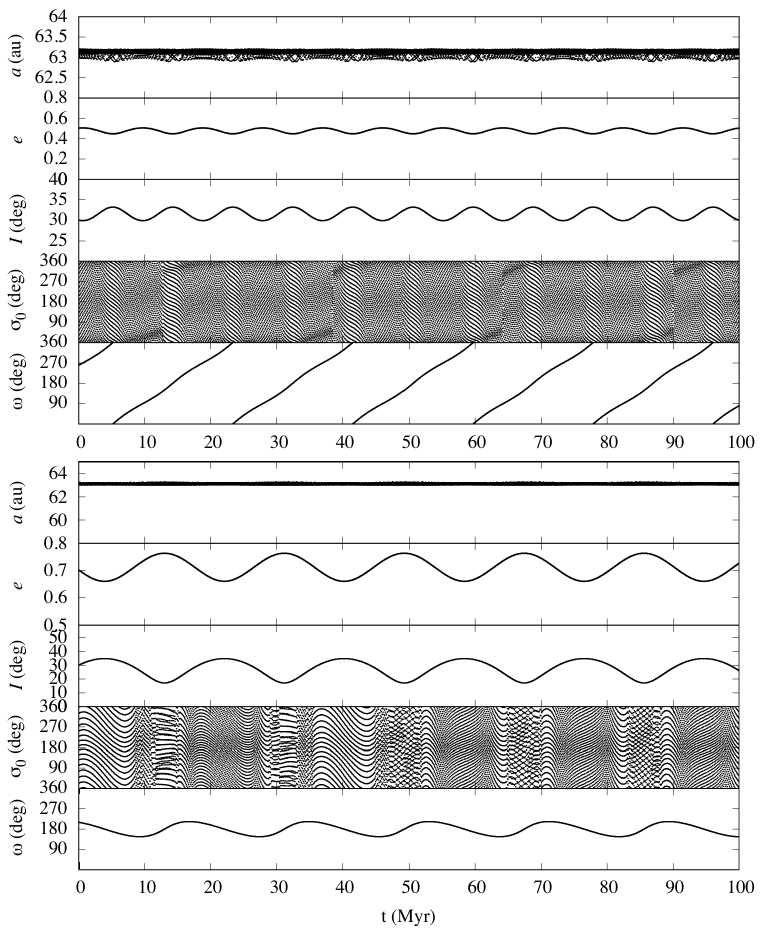}
\end{center}
\caption{Orbital evolution with time for two TNOs near the Kozai resonance over 100~Myr for initial $I=30^\circ$. We show the semi-major axis, the eccentricity, the inclination, the resonant angle $\sigma_0$, and the argument of the pericentre (from top to bottom). The initial conditions correspond to the black dots in Fig.~\ref{stabmap_i30_sigma0_filter_scaled}; i.e. $a=63.1$~au, $e=0.5$, and $\sigma_0=200^\circ$ (top), and $a=63.1$~au, $e=0.7$, and $\sigma_0=300^\circ$ (bottom). \label{kozaitime}}
\end{figure}

In Fig.~\ref{stabmap_i30_sigma0_filter_scaled}, we show the stability maps for $I=30^\circ$ and different eccentricities going from $e \approx 0.1$ to $0.99$.
These maps are to be compared with those obtained for the planar motion ($I=0^\circ$) shown in Fig.~\ref{stabmap_planar_sigma0_filter_scaled}.
Stability maps for $I=60^\circ$ and $I=85^\circ$ are also drawn in Appendix~\ref{appendinc}, but there are no significant differences with respect to the case for $I=30^\circ$.
We nevertheless notice that, for all eccentricity values, the amplitude of the resonant areas decreases as we increase the inclination.
Therefore, capture in the 3/1 MMR resonance becomes more difficult for higher inclination values, at least for the resonant angle $\sigma_0$.

For non-planar systems, we observe that the main resonant structures of the planar case are still present, although some new interesting dynamics also emerge.
For $e \approx 0.1$, in the planar case there was a small single resonant island around $\sigma_0=180^\circ$.
When we increase the inclination, the equilibrium point at $\sigma_0 = 180^\circ$ becomes unstable, and splits into two new resonant areas centred at asymmetric $\sigma_0 = \pm \sigma_r$ values, as happens in the planar case for $e>0.13$.
For $e=0.7$, we observe the opposite behaviour. 
In the planar case, there are three stability regions: two asymmetric regions centred at $\sigma_0 = \pm \sigma_r$ and another symmetric region centred at $\sigma_0 = 0^\circ$.
For $I=30^\circ$, the resonant regions merge in a single wider resonant area centred at $\sigma_0 = 0^\circ$.
A similar behaviour occurs for $e=0.99$. 
In the planar case, there were two resonant areas centred at $\sigma_0 = 0^\circ$ and $180^\circ$, and for $I=30^\circ$ the two regions merge in a single resonant area around $\sigma_0 = 0^\circ$.
Interestingly, for $e=0.9$ the three different resonant regions survive in the inclined case.

Another striking and very interesting feature occurs when we study the dynamics of the inclined TNOs.
In the planar case, all trajectories with $e \ge 0.5$ outside the 3/1 MMR are unstable, because they cross the orbit of Neptune.
However, for $I=30^\circ$, large stability regions are observed outside the resonance.
A closer analysis of these orbits shows that the TNOs are protected by the Kozai resonance, that is, a secular resonance between the precession of the node and the precession of the pericentre, which also prevents close encounters with Neptune.
For trajectories trapped inside the Kozai resonance, the angle $\omega$ can librate around $0^\circ, 90^\circ, 180^\circ$, or $270^\circ$ (Eq.\,(\ref{omegaangle})). 
For trajectories outside this resonance but close to the separatrix, the eccentricity and the inclination can undergo large variations with maxima and minima in phase with the libration centres.
The Kozai dynamics beyond Neptune has been studied in the non-resonant \citep{Gallardo_etal_2012, Saillenfest_etal_2016} and resonant cases \citep{Saillenfest_Lari_2017, Saillenfest_etal_2017b, Lei_etal_2022}, with findings confirming the existence of these different families of libration centres as a function of the eccentricity and inclination of the TNOs.

Due to our choice of initial conditions ($\Omega=0^\circ$), we have $\omega=\varpi=-\sigma_0/2$ for the stability maps, and so libration for $\omega$ around $0^\circ$ and $180^\circ$ occurs for stable regions that cluster around $\sigma_0 = 0^\circ$, while libration around $90^\circ$ and $270^\circ$ occurs for stable regions that cluster around $\sigma_0 = 180^\circ$. 
Fig.~\ref{kozaitime} shows two different examples of trajectories with $a=63.1$~au as a function of time that are stabilised by the Kozai resonance (marked with a black dot in Fig.~\ref{stabmap_i30_sigma0_filter_scaled}), one for $e=0.5$ with $\sigma_0=200^\circ$ ($\omega=260^\circ$), and another for $e=0.7$ with $\sigma_0=300^\circ$ ($\omega=210^\circ$).
In the first example, the TNO is outside the Kozai resonance, but its eccentricity decreases from $0.5$ to nearly $0.4$ whenever $\omega = 0^\circ$ and $\omega = 180^\circ$.
As a result, when the line of nodes between the orbital planes of the TNO and Neptune is aligned with the pericentre, the eccentricity is small enough to prevent close encounters with Neptune (see Fig.~\ref{classification_tno_13mmr}).
In the second example, the TNO is trapped inside the Kozai resonance.
Therefore, although the eccentricity oscillates around 0.7, close encounters with Neptune are expected when $\omega \approx \pm 85^\circ$; however, they never occur because we always have $\omega \approx 180^\circ$. 

\begin{figure}
\begin{center}
\includegraphics[width=0.95\columnwidth]{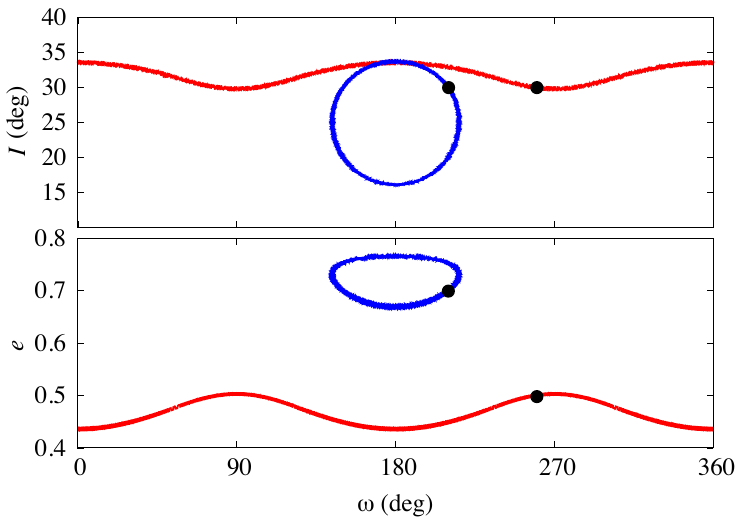}
\end{center}
\caption{Orbital evolution as a function of $\omega$ for two TNOs near the Kozai resonance. We show the inclination (top) and the eccentricity (bottom). The black dots correspond to the initial conditions taken from Figs.~\ref{stabmap_i30_sigma0_filter_scaled} and \ref{kozaitime}, i.e. $a=63.1$~au and $I=30^\circ$, with $e=0.5$ and $\sigma_0=200^\circ$ (in red), and with $e=0.7$ and $\sigma_0=300^\circ$ (in blue).  \label{kozaiomg}}
\end{figure}

For a better understanding of the two kinds of Kozai protection mechanism, Fig.~\ref{kozaiomg} shows the evolution of the eccentricity and inclination as a function of $\omega$ for the previous two trajectories in Fig.~\ref{kozaitime}.
In the case with $e=0.5$ (in red), we observe a clear correlation between the eccentricity minima and the alignment of the line of nodes with the pericentre ($\omega = 0^\circ$ and $180^\circ$). 
In the case with $e=0.7$ (in blue), we observe that $\omega \in [145^\circ,205^\circ]$, and therefore never gets too close to $\omega \approx \pm 85^\circ$.
We also observe a correlation between eccentricity and inclination.
As the semi-major axis is approximately constant (Fig.~\ref{kozaitime}), these variations are given by the `Kozai integral', that is, $\sqrt{1-e^2} \cos I \approx const$ (Eq.\,(\ref{N_nonplanar})).

\begin{figure*}
\includegraphics[width=1.0\textwidth]{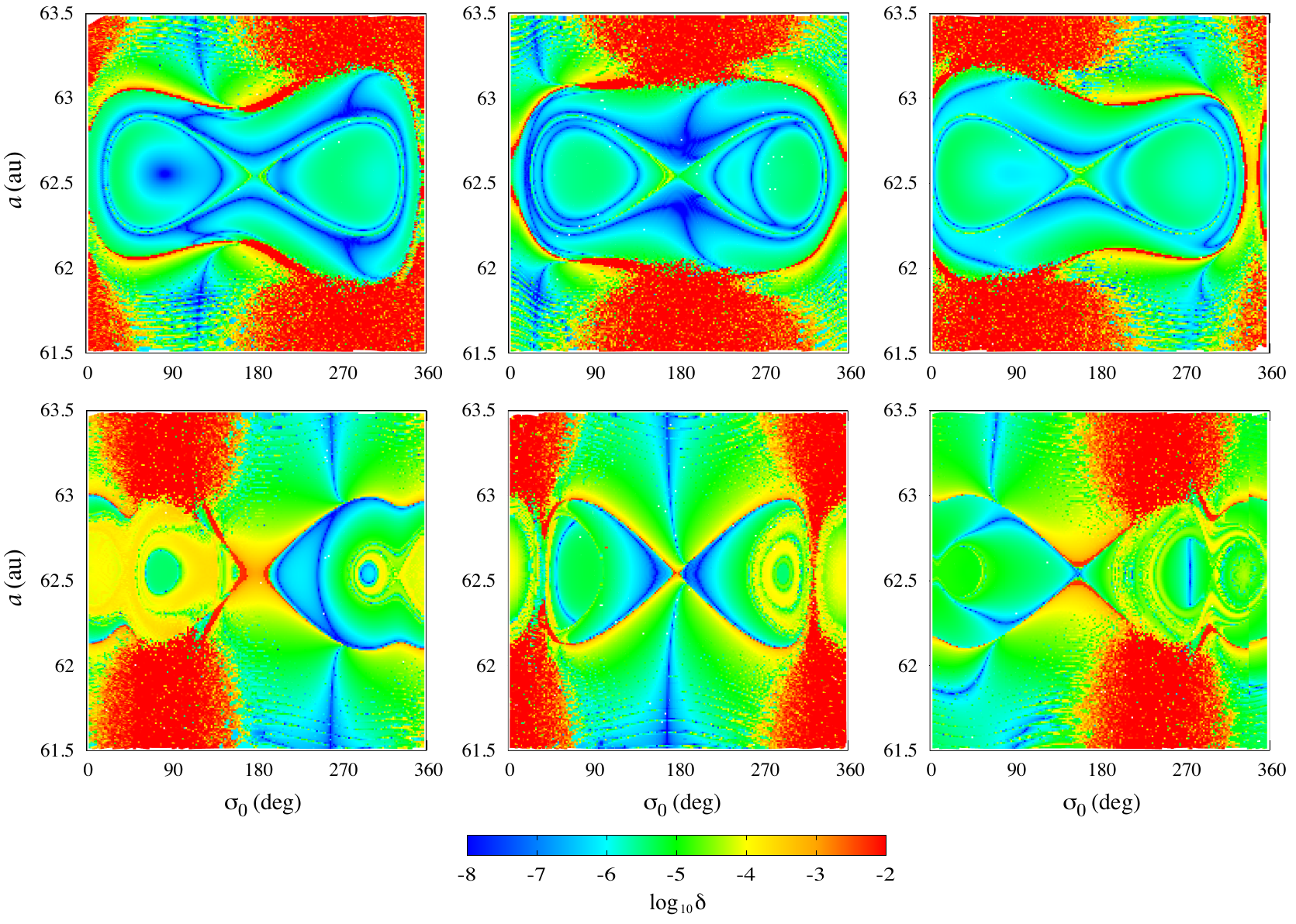}
      \caption{Stability maps in the plane ($\sigma_0, a$) for the three-body restricted problem close to the 3/1 MMR with $I=30^\circ$ and $\Omega=45^\circ$, $90^\circ$, and $135^\circ$ from left to right. For each panel, $N$ is constant and parametrised by the initial eccentricities $e=0.5$ (top) and  $e=0.7$ (bottom). The colour scale corresponds to the relative frequency diffusion index in logarithmic scale (Eq.\,(\ref{deltaindex})). \label{stabmap_i30_sigma0_filter_scaled_Omega}}
\end{figure*}

\begin{figure*}
\includegraphics[width=1.0\textwidth]{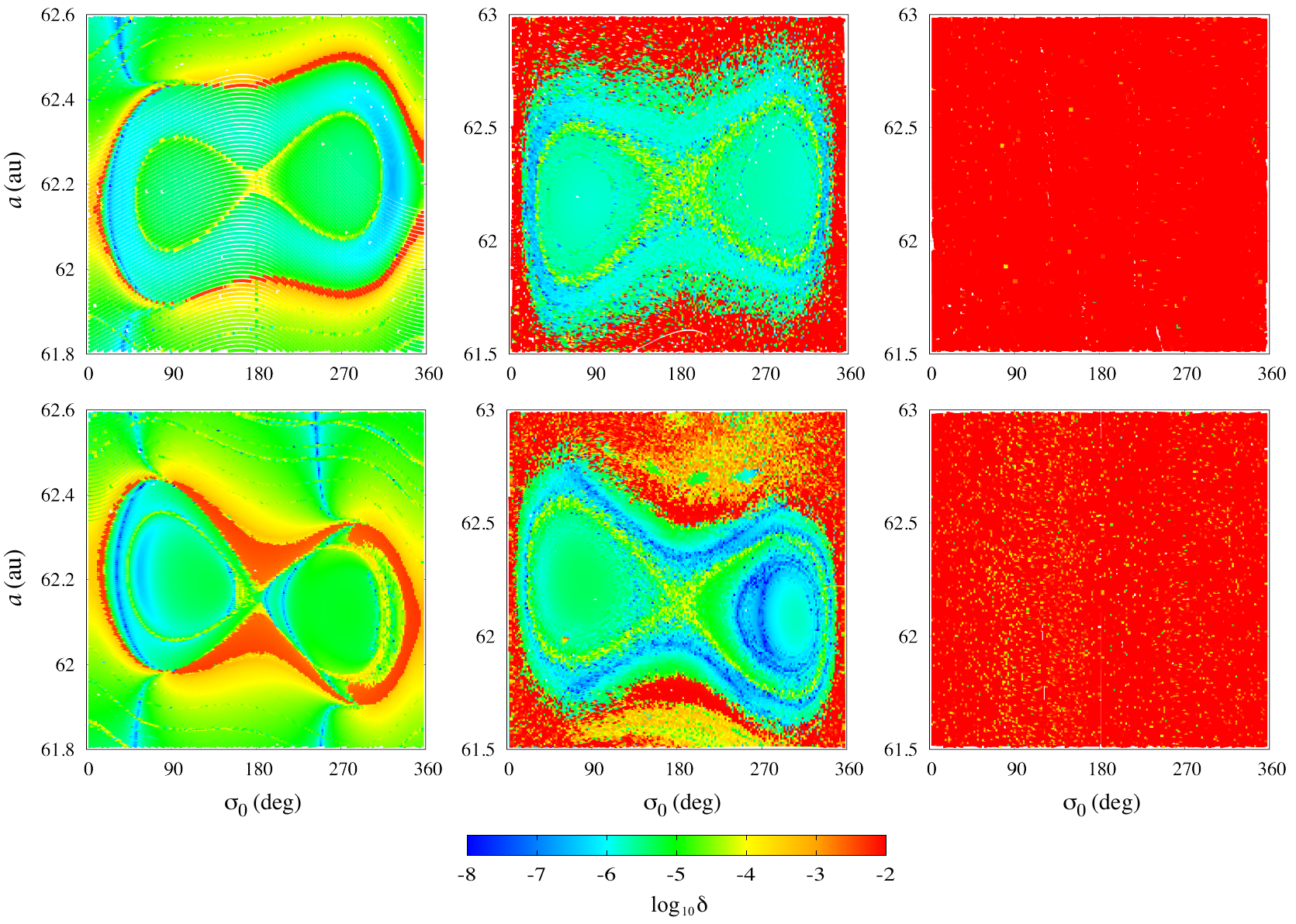}
\caption{Stability maps in the plane ($\sigma_0, a$) close to the 3/1 MMR in the presence of the four giant planets with $I=0^\circ$ (top) and $I=30^\circ$ (bottom). Each panel is parametrised by a different initial eccentricity, $e=0.3$ (left), $e=0.5$ (middle), and  $e=0.7$ (right). The colour scale corresponds to the relative frequency diffusion index in logarithmic scale (Eq.\,(\ref{deltaindex})). \label{stabmap_multiplanetary_planar}}
\end{figure*}

The position of the Kozai resonance depends on the angle $\Omega$ (Eq.\,(\ref{omegaangle})), whose initial value was fixed at zero in Fig.~\ref{stabmap_i30_sigma0_filter_scaled}.
This choice should not impact the 3/1 MMR, as $\sigma_0$ does not depend on $\Omega$ (Eq.\,(\ref{resonantangle})), but it will impact the location of the Kozai regions as a function of $\sigma_0$, because when $\Omega \ne 0$ we have $\omega=-\sigma_0/2-\Omega$.
Therefore, in Fig.~\ref{stabmap_i30_sigma0_filter_scaled_Omega} we plot the stability maps for $I=30^\circ$ and for the eccentricities $e\approx0.5$ and 0.7, but now using\footnote{As $\sigma_0=-2(\omega+\Omega)$, the maps are identical for $\Omega$ and $\Omega+180^\circ$.} $\Omega = 45^\circ$, $90^\circ$ and $135^\circ$.
As expected, the 3/1 MMR islands are still present at the same locations, but the Kozai stable regions are shifted by $- 2\Omega$.

We additionally observe that the 3/1 MMR dynamics cannot be completely dissociated from the Kozai dynamics.
For $e\approx 0.5$, the Kozai areas seem to `compress' the size of the libration width in the 3/1 MMR, at least for the outer regions, which encircle both the two asymmetric $\pm\sigma_r$ equilibria.
For $e\approx 0.7$, the interaction dynamics is even richer.
While for $\Omega=0^\circ$ the 3/1 MMR asymmetric islands centred at $\pm\sigma_r$ were merged with the symmetric island centred at $\sigma_0=0^\circ$, for $\Omega=90^\circ$ we return to the configuration with three resonant centres, as in the planar case.
For the intermediate $\Omega = 45^\circ$ and $\Omega = 135^\circ$, the unstable regions overlap only one side of the asymmetric resonant islands, which becomes much more unstable than the side `protected' by the Kozai mechanism.
As a result, the asymmetric island within the Kozai region merges with the symmetric island, while the other asymmetric island remains isolated.

\subsection{Perturbations by the giant planets}
\llabel{otherplanets}

The stability maps also allow us to easily explore the dynamics in the presence of the remaining planets.
As in previous studies, we take into account only the gravitational interactions with the giant planets (Jupiter, Saturn, Uranus and Neptune), because the perturbations from the rocky planets are not particularly important \citep[e.g.][]{Kuchner_etal_2002, Saillenfest_Lari_2017}.
For simplicity, we place the four planets initially in circular orbits and in the same reference plane ($I'=0$), and all the initial longitudes are taken equal to zero.
We explore the dynamics of the TNO in the plane ($\sigma_0, a$) and obtain the initial orbital elements in the same way as described in Sect.~\ref{nonplanardyn}.

In Fig.~\ref{stabmap_multiplanetary_planar} (top), we show the stability maps in the coplanar case ($I=0^\circ$) and for different eccentricities, $e = 0.3$, $0.5$, and $0.7$.
These maps are to be compared with those obtained for the three-body problem shown in Fig.~\ref{stabmap_planar_sigma0_filter_scaled}.
For $e=0.3$ and $0.5$, the dynamics around the 3/1 MMR remains mostly unchanged;
we only note an increase in the diffusion index levels, which is normal, because the mean motion of the TNO undergoes additional perturbations from the remaining planets.
For $e=0.3$, the non-resonant regions remain stable, but they become unstable for $e=0.5$ because of close encounters with Neptune, as was found to be the case for the three-body problem (Fig.~\ref{stabmap_planar_sigma0_filter_scaled}).
However, for $e=0.7$, we observe that all trajectories become unstable, including those that were trapped in the 3/1 MMR, because these TNOs now cross the orbit of Uranus.
Indeed, in the frequency analysis presented in Sect.~\ref{freqanalys}, the four TNOs presently observed in this region (see dashed line in Fig.~\ref{classification_tno_13mmr}) are completely unstable and are destroyed within a few million years.
These objects merit further investigation in order to understand how they evolved into the present position, but this is out of the scope of the present paper.

In Fig.~\ref{stabmap_multiplanetary_planar} (bottom), we show the stability maps for $I=30^\circ$ and for the same previous eccentricities, $e = 0.3$, $0.5$, and $0.7$.
These maps are to be compared with those obtained for the three-body problem shown in Fig.~\ref{stabmap_i30_sigma0_filter_scaled}.
For $e=0.3$, the dynamics around the 3/1 MMR is also similar to the three-body problem, apart from the increase in the diffusion index levels.
However, for $e=0.5$, there is an important difference: the 3/1 MMR is still present, but the non-resonant stable areas now disappear as in the planar case.
The stability achieved outside the 3/1 MMR in the three-body problem for $e \ge 0.5$ is possible due to the presence of the Kozai resonance (Sect.~\ref{nonplanardyn}). 
When we include the remaining planets, they introduce more secular frequencies that modify the precession of the node and the precession of the pericentre of the TNO.
As a result, the corresponding longitudes no longer evolve at the same pace and the argument of the pericentre (Eq.\,(\ref{omegaangle})) no longer librates or is in phase with the eccentricity oscillations, which leads to a subsequent close encounter with Neptune and ejection of the TNO.
For $e=0.7$, as in the planar case, we observe that all trajectories become unstable, because they cross the orbit of Uranus.

When we compare the planar dynamics (Fig.~\ref{stabmap_multiplanetary_planar}~top) with the dynamics at $I=30^\circ$ (Fig.~\ref{stabmap_multiplanetary_planar}~bottom) in the presence of the four giant planets, we conclude that, as opposed to the three-body problem, there are no significant differences.
The key ingredients to assure stable orbits for the {threetinos} family is to be captured in the 3/1 MMR with $e \lesssim 0.5$ (higher eccentricities are scattered by Uranus).
Orbits with $e \lesssim 0.4$ can also be stable outside the 3/1 MMR, because they no longer cross the orbit of Neptune.
These results are in perfect agreement with most of the currently observed TNOs in the area (Fig.~\ref{classification_tno_13mmr}).

\section{Conclusions}

\label{conclsec}

A large number of TNOs are observed near a 3/1 MMR with Neptune, the majority of them with significant inclinations and eccentricities between 0.3 and 0.5 (Fig.~\ref{classification_tno_13mmr}).
In this paper, we present a study of the dynamics of these objects.
We first selected those that can be trapped in exact resonance and identify the main libration angle, $\sigma_0$, using frequency analysis. 

We initially built a secular planar model that can be reduced to a problem with a single degree of freedom.
We performed a canonical change of variables to use $\sigma_0$ as the free coordinate.
This allows us to follow the evolution of the centres of the equilibria with eccentricity and to understand the main dynamical structures in the problem.
We then studied how the secular model undergoes modification in the non-averaged planar case using surface sections, but also using stability maps. 
We show that stability maps are able to capture all the dynamical structures of the problem and additionally provide a clear view of the chaotic diffusion and unstable zones.
Therefore, we finally use these maps to study the non-planar dynamics and in the presence of perturbations from the remaining giant planets.

In the frame of the three-body problem, we observe that when we increase the inclination, the dynamics in the 3/1 MMR does not change significantly with respect to the planar case.
There are nevertheless some resonant islands that split (for $e=0.1$) or merge (for $e=0.7$).
The amplitude of the libration areas around $\sigma_0$ also decreases as we increase the inclination.
Therefore, capture in the 3/1 MMR for the resonant angle $\sigma_0$ becomes more difficult for higher inclination values.
The most striking difference concerns the secular dynamics. 
In the planar case, all TNOs with $e \ge 0.5$ outside the 3/1 MMR are unstable, because they cross the orbit of Neptune.
In the inclined case, large stability regions are observed, which correspond to trajectories trapped in a Kozai secular resonance.
The presence of the Kozai resonance also renders the dynamics in 3/1 MMR more stable, leading to a merger of the libration regions of previously isolated resonant areas.

Using stability maps, we finally show that the inclusion of the remaining giant planets in the model also does not significantly change the global dynamics of the 3/1 MMR.
The most important difference is that TNOs with $e > 0.5$ cannot survive, as these trajectories cross the orbit of Uranus.
In addition, the Kozai stable areas disappear even for TNOs with $e \lesssim 0.5$, because the perturbations from the planets modify the precession rates of the node and the pericentre of the TNO.
Indeed, we confirm that the orbits of the three TNOs observed in this region (Fig.~\ref{classification_tno_13mmr}) are completely unstable and are destroyed in a few million years.

We focused the present study on the 3/1 MMR with Neptune, but our approach is completely general and can therefore be extended to other MMRs beyond Neptune or to the asteroid belt.
Stability maps have already been used to study the dynamics of test particles in the Solar System near MMR \citep[e.g.][]{Robutel_Laskar_2001, Gallardo_2019, Forgacs-Dajka_etal_2023, Giuppone_etal_2023}. 
However, these maps are usually plotted in the plane $(a,e)$, which only provides the location of the resonances and their width as a function of eccentricity.
We show here that a complete understanding of the full dynamics near MMR is only possible when we additionally plot these maps as a function of the resonant angle. 
Indeed, complex structures such as multiple resonant islands and Kozai interactions only become clear in this projection.

\begin{acknowledgements}
We thank Helena Morais for discussions. 
We are grateful to the referee Jean-Marc Petit for his insightful comments.
This work was supported by
CFisUC strategic project (UIDB/04564/2020 and UIDP/04564/2020),
GRAVITY (PTDC/FIS-AST/7002/2020),
SmartGlow (POCI-01-0247-FEDER-069733),
PHOBOS (POCI-01-0145-FEDER-029932), and
ENGAGE SKA (POCI-01-0145-FEDER-022217),
funded by COMPETE 2020 and FCT, Portugal.
We acknowledge the Laboratory for Advanced Computing at University of Coimbra (\href{https://www.uc.pt/lca}{https://www.uc.pt/lca}) for providing HPC resources to perform the stability maps with high resolution.
This research has made use of data provided by the International Astronomical Union's Minor Planet Center.
\end{acknowledgements}

\bibliographystyle{aa}
\bibliography{\bibpath correia}

\begin{thebibliography}{46}
\expandafter\ifx\csname natexlab\endcsname\relax\def\natexlab#1{#1}\fi

\bibitem[{{Bannister} {et~al.}(2018){Bannister}, {Gladman}, {Kavelaars},
  {Petit}, {Volk}, {Chen}, {Alexandersen}, {Gwyn}, {Schwamb}, {Ashton},
  {Benecchi}, {Cabral}, {Dawson}, {Delsanti}, {Fraser}, {Granvik},
  {Greenstreet}, {Guilbert-Lepoutre}, {Ip}, {Jakubik}, {Jones}, {Kaib},
  {Lacerda}, {Van Laerhoven}, {Lawler}, {Lehner}, {Lin}, {Lykawka}, {Marsset},
  {Murray-Clay}, {Pike}, {Rousselot}, {Shankman}, {Thirouin}, {Vernazza}, \&
  {Wang}}]{Bannister_etal_2018}
{Bannister}, M.~T., {Gladman}, B.~J., {Kavelaars}, J.~J., {et~al.} 2018, \apjs,
  236, 18

\bibitem[{{Beauge}(1994)}]{Beauge_1994}
{Beauge}, C. 1994, Celestial Mechanics and Dynamical Astronomy, 60, 225

\bibitem[{{Beaug{\'e}} {et~al.}(2006){Beaug{\'e}}, {Michtchenko}, \&
  {Ferraz-Mello}}]{Beauge_etal_2006}
{Beaug{\'e}}, C., {Michtchenko}, T.~A., \& {Ferraz-Mello}, S. 2006, \mnras,
  365, 1160

\bibitem[{{Carpino} {et~al.}(1987){Carpino}, {Milani}, \&
  {Nobili}}]{Carpino_etal_1987}
{Carpino}, M., {Milani}, A., \& {Nobili}, A.~M. 1987, \aap, 181, 182

\bibitem[{{Couetdic} {et~al.}(2010){Couetdic}, {Laskar}, {Correia}, {Mayor}, \&
  {Udry}}]{Couetdic_etal_2010}
{Couetdic}, J., {Laskar}, J., {Correia}, A.~C.~M., {Mayor}, M., \& {Udry}, S.
  2010, \aap, 519, A10

\bibitem[{{Crompvoets} {et~al.}(2022){Crompvoets}, {Lawler}, {Volk}, {Chen},
  {Gladman}, {Peltier}, {Alexandersen}, {Bannister}, {Gwyn}, {Kavelaars}, \&
  {Petit}}]{Crompvoets_etal_2022}
{Crompvoets}, B.~L., {Lawler}, S.~M., {Volk}, K., {et~al.} 2022, \psj, 3, 113

\bibitem[{{Dumas} \& {Laskar}(1993)}]{Dumas_Laskar_1993}
{Dumas}, H.~S. \& {Laskar}, J. 1993, \prl, 70, 2975

\bibitem[{{Duncan} {et~al.}(1995){Duncan}, {Levison}, \&
  {Budd}}]{Duncan_etal_1995}
{Duncan}, M.~J., {Levison}, H.~F., \& {Budd}, S.~M. 1995, \aj, 110, 3073

\bibitem[{{Elliot} {et~al.}(2005){Elliot}, {Kern}, {Clancy}, {Gulbis},
  {Millis}, {Buie}, {Wasserman}, {Chiang}, {Jordan}, {Trilling}, \&
  {Meech}}]{Elliot_etal_2005}
{Elliot}, J.~L., {Kern}, S.~D., {Clancy}, K.~B., {et~al.} 2005, \aj, 129, 1117

\bibitem[{{Fern{\'a}ndez} {et~al.}(2004){Fern{\'a}ndez}, {Gallardo}, \&
  {Brunini}}]{Fernandez_etal_2004}
{Fern{\'a}ndez}, J.~A., {Gallardo}, T., \& {Brunini}, A. 2004, \icarus, 172,
  372

\bibitem[{{Fernandez} \& {Ip}(1984)}]{Fernandez_Ip_1984}
{Fernandez}, J.~A. \& {Ip}, W.~H. 1984, \icarus, 58, 109

\bibitem[{{Forg{\'a}cs-Dajka} {et~al.}(2023){Forg{\'a}cs-Dajka}, {Kov{\'a}ri},
  {Kov{\'a}cs}, {Kiss}, \& {S{\'a}ndor}}]{Forgacs-Dajka_etal_2023}
{Forg{\'a}cs-Dajka}, E., {Kov{\'a}ri}, E., {Kov{\'a}cs}, T., {Kiss}, C., \&
  {S{\'a}ndor}, Z. 2023, \apjs, 266, 5

\bibitem[{{Gallardo}(2019)}]{Gallardo_2019}
{Gallardo}, T. 2019, \icarus, 317, 121

\bibitem[{{Gallardo} {et~al.}(2012){Gallardo}, {Hugo}, \&
  {Pais}}]{Gallardo_etal_2012}
{Gallardo}, T., {Hugo}, G., \& {Pais}, P. 2012, \icarus, 220, 392

\bibitem[{Gastineau \& Laskar(2011)}]{Gastineau_Laskar_2011}
Gastineau, M. \& Laskar, J. 2011, ACM Commun. Comput. Algebra, 44, 194

\bibitem[{{Giuppone} {et~al.}(2023){Giuppone}, {Rodr{\'\i}guez}, {Alencastro},
  {Roig}, \& {Gallardo}}]{Giuppone_etal_2023}
{Giuppone}, C., {Rodr{\'\i}guez}, A., {Alencastro}, V., {Roig}, F., \&
  {Gallardo}, T. 2023, Celestial Mechanics and Dynamical Astronomy, 135, 3

\bibitem[{{Gladman} {et~al.}(2002){Gladman}, {Holman}, {Grav}, {Kavelaars},
  {Nicholson}, {Aksnes}, \& {Petit}}]{Gladman_etal_2002}
{Gladman}, B., {Holman}, M., {Grav}, T., {et~al.} 2002, \icarus, 157, 269

\bibitem[{{Gladman} {et~al.}(2008){Gladman}, {Marsden}, \&
  {Vanlaerhoven}}]{Gladman_etal_2008}
{Gladman}, B., {Marsden}, B.~G., \& {Vanlaerhoven}, C. 2008, in The Solar
  System Beyond Neptune, ed. M.~A. {Barucci}, H.~{Boehnhardt}, D.~P.
  {Cruikshank}, A.~{Morbidelli}, \& R.~{Dotson}, 43

\bibitem[{{Goldstein}(1950)}]{Goldstein_1950}
{Goldstein}, H. 1950, {Classical mechanics} (Addison-Wesley, Reading)

\bibitem[{{Gomes} {et~al.}(2005){Gomes}, {Levison}, {Tsiganis}, \&
  {Morbidelli}}]{Gomes_etal_2005}
{Gomes}, R., {Levison}, H.~F., {Tsiganis}, K., \& {Morbidelli}, A. 2005, \nat,
  435, 466

\bibitem[{{Hadjidemetriou}(1993)}]{Hadjidemetriou_1993}
{Hadjidemetriou}, J.~D. 1993, Celestial Mechanics and Dynamical Astronomy, 56,
  201

\bibitem[{{Hahn} \& {Malhotra}(2005)}]{Hahn_Malhotra_2005}
{Hahn}, J.~M. \& {Malhotra}, R. 2005, \aj, 130, 2392

\bibitem[{{Ichtiaroglou} {et~al.}(1989){Ichtiaroglou}, {Katopodis}, \&
  {Michalodimitrakis}}]{Ichtiaroglou_etal_1989}
{Ichtiaroglou}, S., {Katopodis}, K., \& {Michalodimitrakis}, M. 1989, Journal
  of Astrophysics and Astronomy, 10, 367

\bibitem[{{Jewitt}(2005)}]{Jewitt_2005}
{Jewitt}, D. 2005, \aj, 129, 530

\bibitem[{{Kuchner} {et~al.}(2002){Kuchner}, {Brown}, \&
  {Holman}}]{Kuchner_etal_2002}
{Kuchner}, M.~J., {Brown}, M.~E., \& {Holman}, M. 2002, \aj, 124, 1221

\bibitem[{{Lan} \& {Malhotra}(2019)}]{Lan_Malhotra_2019}
{Lan}, L. \& {Malhotra}, R. 2019, Celestial Mechanics and Dynamical Astronomy,
  131, 39

\bibitem[{{Laskar}(1990)}]{Laskar_1990}
{Laskar}, J. 1990, \icarus, 88, 266

\bibitem[{{Laskar}(1993)}]{Laskar_1993PD}
{Laskar}, J. 1993, Physica D Nonlinear Phenomena, 67, 257

\bibitem[{{Laskar} \& {Robutel}(2001)}]{Laskar_Robutel_2001}
{Laskar}, J. \& {Robutel}, P. 2001, Celestial Mechanics and Dynamical
  Astronomy, 80, 39

\bibitem[{{Lawler} {et~al.}(2019){Lawler}, {Pike}, {Kaib}, {Alexandersen},
  {Bannister}, {Chen}, {Gladman}, {Gwyn}, {Kavelaars}, {Petit}, \&
  {Volk}}]{Lawler_etal_2019}
{Lawler}, S.~M., {Pike}, R.~E., {Kaib}, N., {et~al.} 2019, \aj, 157, 253

\bibitem[{{Lei} {et~al.}(2022){Lei}, {Li}, {Huang}, \& {Li}}]{Lei_etal_2022}
{Lei}, H., {Li}, J., {Huang}, X., \& {Li}, M. 2022, arXiv e-prints,
  arXiv:2207.12954

\bibitem[{{Malhotra} {et~al.}(2018){Malhotra}, {Lan}, {Volk}, \&
  {Wang}}]{Malhotra_etal_2018}
{Malhotra}, R., {Lan}, L., {Volk}, K., \& {Wang}, X. 2018, \aj, 156, 55

\bibitem[{{Message}(1958)}]{Message_1958}
{Message}, P.~J. 1958, \aj, 63, 443

\bibitem[{{Morbidelli} {et~al.}(2007){Morbidelli}, {Tsiganis}, {Crida},
  {Levison}, \& {Gomes}}]{Morbidelli_etal_2007}
{Morbidelli}, A., {Tsiganis}, K., {Crida}, A., {Levison}, H.~F., \& {Gomes}, R.
  2007, \aj, 134, 1790

\bibitem[{{Namouni} \& {Morais}(2018)}]{Namouni_Morais_2018b}
{Namouni}, F. \& {Morais}, M.~H.~M. 2018, \mnras, 474, 157

\bibitem[{{Namouni} \& {Morais}(2020)}]{Namouni_Morais_2020b}
{Namouni}, F. \& {Morais}, M.~H.~M. 2020, \mnras, 493, 2854

\bibitem[{{Pichierri} {et~al.}(2017){Pichierri}, {Morbidelli}, \&
  {Lai}}]{Pichierri_etal_2017}
{Pichierri}, G., {Morbidelli}, A., \& {Lai}, D. 2017, \aap, 605, A23

\bibitem[{{Robutel} \& {Laskar}(2001)}]{Robutel_Laskar_2001}
{Robutel}, P. \& {Laskar}, J. 2001, \icarus, 152, 4

\bibitem[{{Saillenfest} {et~al.}(2016){Saillenfest}, {Fouchard}, {Tommei}, \&
  {Valsecchi}}]{Saillenfest_etal_2016}
{Saillenfest}, M., {Fouchard}, M., {Tommei}, G., \& {Valsecchi}, G.~B. 2016,
  Celestial Mechanics and Dynamical Astronomy, 126, 369

\bibitem[{{Saillenfest} {et~al.}(2017){Saillenfest}, {Fouchard}, {Tommei}, \&
  {Valsecchi}}]{Saillenfest_etal_2017b}
{Saillenfest}, M., {Fouchard}, M., {Tommei}, G., \& {Valsecchi}, G.~B. 2017,
  Celestial Mechanics and Dynamical Astronomy, 127, 477

\bibitem[{{Saillenfest} \& {Lari}(2017)}]{Saillenfest_Lari_2017}
{Saillenfest}, M. \& {Lari}, G. 2017, \aap, 603, A79

\bibitem[{{Taylor}(1983{\natexlab{a}})}]{Taylor_1983a}
{Taylor}, D.~B. 1983{\natexlab{a}}, Celestial Mechanics, 29, 75

\bibitem[{{Taylor}(1983{\natexlab{b}})}]{Taylor_1983b}
{Taylor}, D.~B. 1983{\natexlab{b}}, Celestial Mechanics, 29, 51

\bibitem[{{Tsiganis} {et~al.}(2005){Tsiganis}, {Gomes}, {Morbidelli}, \&
  {Levison}}]{Tsiganis_etal_2005}
{Tsiganis}, K., {Gomes}, R., {Morbidelli}, A., \& {Levison}, H.~F. 2005, \nat,
  435, 459

\bibitem[{{Voyatzis} \& {Kotoulas}(2005)}]{Voyatzis_Kotoulas_2005}
{Voyatzis}, G. \& {Kotoulas}, T. 2005, \planss, 53, 1189

\bibitem[{{Voyatzis} {et~al.}(2018){Voyatzis}, {Tsiganis}, \&
  {Antoniadou}}]{Voyatzis_etal_2018}
{Voyatzis}, G., {Tsiganis}, K., \& {Antoniadou}, K.~I. 2018, Celestial
  Mechanics and Dynamical Astronomy, 130, 29

\end{thebibliography}

\clearpage
\onecolumn

\begin{appendix}

\section{Stability maps for inclined TNOs}

\llabel{appendinc}

Here we provide the stability maps for $I=60^\circ$ and $I=85^\circ$ and different eccentricities going from $e \approx 0.1$ to $0.99$  (we only consider the three-body problem).
These maps are to be compared with those obtained for planar motion ($I=0^\circ$) shown in Fig.~\ref{stabmap_planar_sigma0_filter_scaled} and for $I=30^\circ$ shown in Fig.~\ref{stabmap_i30_sigma0_filter_scaled}.
As we increase the inclination, we observe that the global dynamics does not change much, but the amplitude of the $\sigma_0$ libration area decreases.

\begin{figure*}[h!]
\includegraphics[width=1.0\textwidth]{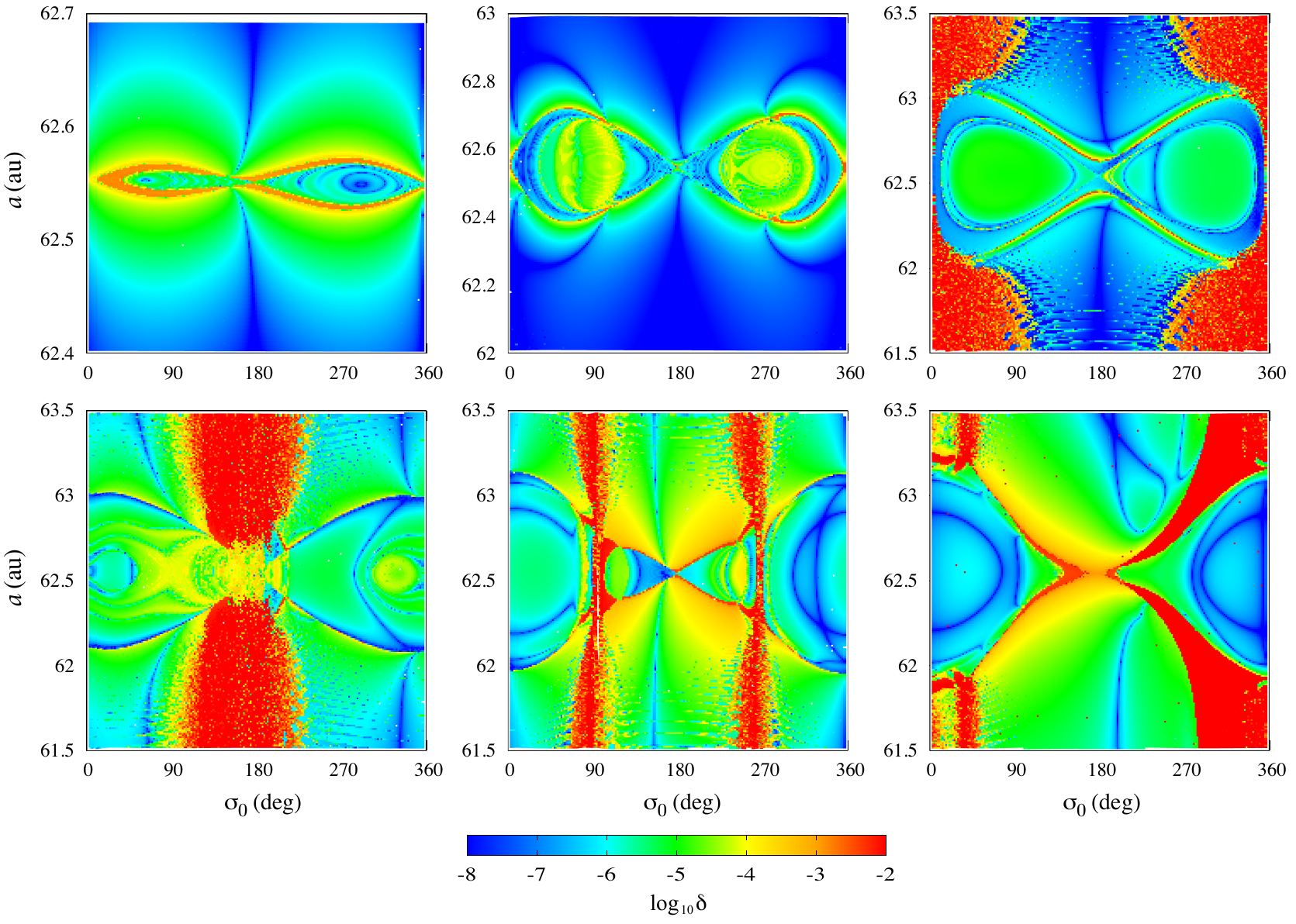}
      \caption{Stability maps in the plane ($\sigma_0, a$) for the three-body restricted problem close to the 3/1 MMR with $I=60^\circ$ and $\Omega=0^\circ$. For each panel, $N$ is constant and parametrised by the initial eccentricities (from left to right):  $e=0.1$, $e=0.3$,  $e=0.5$ (top), and  $e=0.7$, $0.9$, $0.99$ (bottom). The colour scale corresponds to the relative frequency diffusion index in logarithmic scale (Eq.\,(\ref{deltaindex})). \label{stabmap_i60_sigma0_filter_scaled}}
\end{figure*}

\begin{figure*}[h!]
\includegraphics[width=1.0\textwidth]{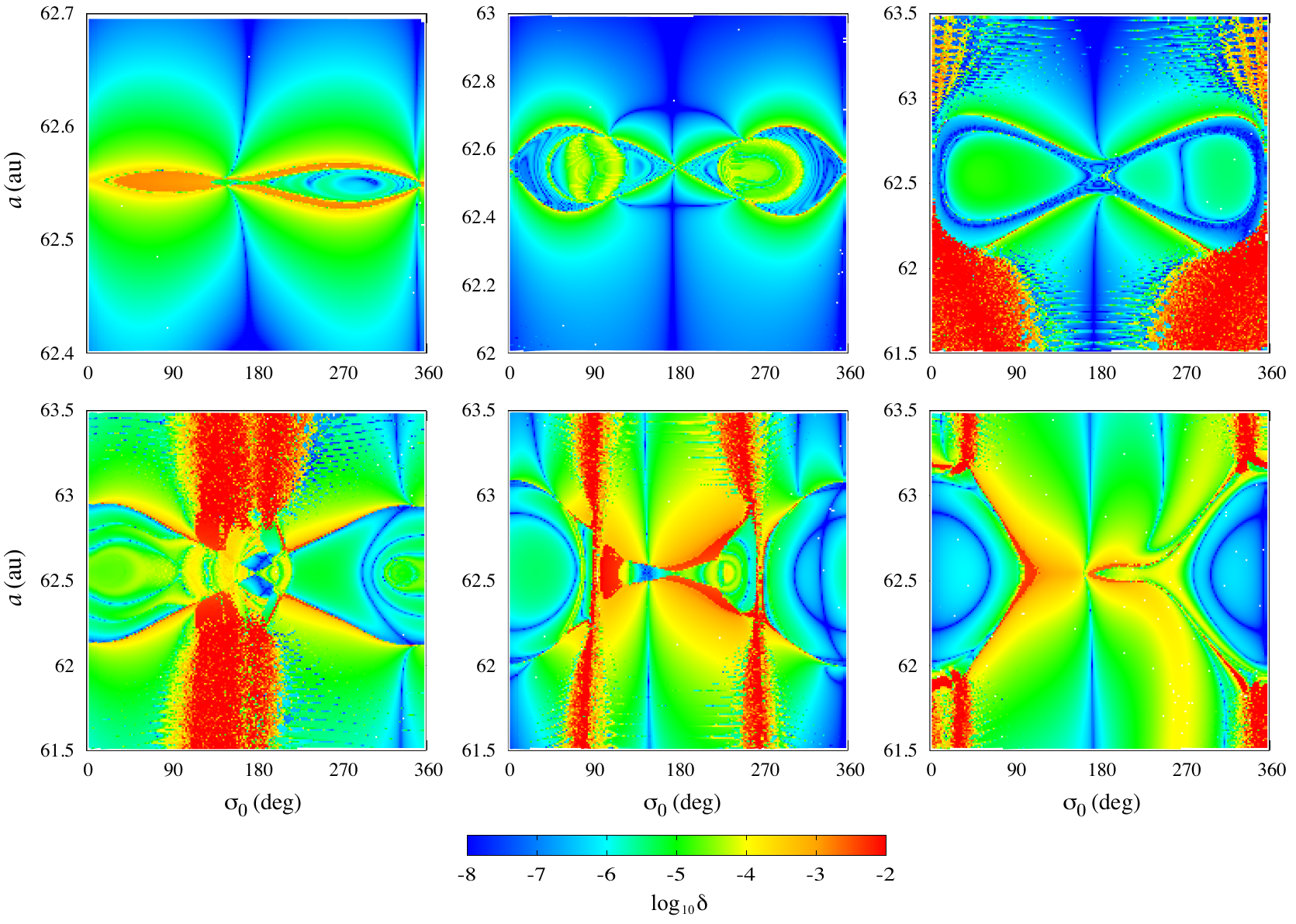}
      \caption{Stability maps in the plane ($\sigma_0, a$) for the three-body restricted problem close to the 3/1 MMR with $I=85^\circ$ and $\Omega=0^\circ$. For each panel, $N$ is constant and parametrised by the initial eccentricities (from left to right):  $e=0.1$, $e=0.3$,  $e=0.5$ (top), and  $e=0.7$, $0.9$, $0.99$ (bottom). The colour scale corresponds to the relative frequency diffusion index in logarithmic scale (Eq.\,(\ref{deltaindex})). \label{stabmap_i90_sigma0_filter_scaled}}
\end{figure*}

\end{appendix}

\end{document}